\documentclass[prb,twocolumn,showpacs,preprintnumbers,amsmath,amssymb]{revtex4-1}

\usepackage{graphicx}

\begin{document}
\title{ Frustrated superfluids in a non-Abelian flux }
\author{ Fadi Sun$^{1,2,4}$, Junsen Wang $^{1,3}$, Jinwu Ye $^{1,2,4}$ and  Youjin Deng $^{3}$  }
\affiliation{
$^{1}$ Department of Physics and Astronomy, Mississippi State University, MS, 39762, USA \\
$^{2}$ Department of Physics, Capital Normal University,
Key Laboratory of Terahertz Optoelectronics, Ministry of Education, and Beijing Advanced innovation Center for Imaging Technology,
Beijing, 100048, China   \\
$^{3}$ CAS Center for Excellence and Synergetic Innovation Center in Quantum Information and Quantum Physics,
University of Science and Technology of China, Hefei, Anhui 230026, China  \\
$^{4}$ Kavli Institute of Theoretical Physics, University of California, Santa Barbara, Santa Barbara, CA 93106  }
\date{\today }

\begin{abstract}
    We study possible superfluid states of the Rashba spin-orbit coupled (SOC)
    spinor bosons with the spin anisotropic interaction $ \lambda $ hopping in a square lattice.
    The frustrations from the  non-abelian flux due to the SOC  leads to novel spin-bond correlated superfluids.
    By using a recently developed systematic "order from quantum disorder" analysis, we not only determine
    the true quantum ground state, but also evaluate the mass gap in the spin sector at $ \lambda < 1 $, especially compute the
    the excitation spectrum of the Goldstone mode in the spin sector at $ \lambda=1 $ which would be quadratic without the  analysis.
    The analysis also leads to different critical exponents on the two sides of the 2nd order transition
    driven by a roton touchdown at $ \lambda=1 $.
    The intimate analogy at $ \lambda=1 $  with the charge neutral Goldstone mode
    in the pseudo-spin sector in the Bilayer quantum Hall systems at the total filling factor $ \nu_T=1 $ are stressed.
    The experimental implications and detections of these novel phenomena in cold atoms loaded on a optical lattice are presented.
\end{abstract}

\maketitle

 There were previous extensive research efforts to study the geometric frustrations on
 the boson superfluids in a triangular \cite{gan,ss1,ss2,ss3,tri,trinnn,five,yan}
 and a Kagome lattice \cite{demler,ka1,ka2,ka3,yan}.  There were also more recent investigations
 on the Abelian flux induced frustrations to the boson superfluids in a bipartite lattice \cite{abe1,abe2,abesf}.
 However, the interplay of Rashba or Dresselhaus spin-orbit coupling (SOC) with interactions
 has not been studied in any bosonic superfluid systems yet. The SOC leads to
 a non-Abelian flux which can be characterized by various Wilson loops \cite{rh}.

 The Rashba or Dresselhaus spin-orbit coupling (SOC) \cite{rashba,ahe,socsemi,ahe2,she,niu,aherev,sherev} is ubiquitous in various
 2d or layered non-centrosymmetric  insulators, semi-conductor systems, metals and superconductors.
 The lattice regularization of any linear combination of the Rashba and Dresselhaus SOC
 $ \alpha k_x \sigma_x + \beta k_y \sigma_y $ is the kinetic term in Eq.\ref{intlambda}.
 The anisotropy can be adjusted by the strains, the shape of the surface or gate electric fields.
 In the cold atom side, there were experimental advances to generate 2d Rashba SOC for
 the fermion $ ^{40} K $ gas \cite{expk40,expk40zeeman} and the tunable quantum anomalous Hall (QAH) SOC
 for $ ^{87} Rb $ atoms in a square lattice\cite{2dsocbec}.
 More recently, the optical lattice clock scheme has been implemented to generate very long lifetime 1d SOC for various atoms
 \cite{SD,clock,clock1,clock2,SDRb,ben}. The 2d Rashba SOC  with a long enough lifetime in a square lattice maybe generated in some near future cold atom experiments.

\begin{figure}
\includegraphics[width=2cm]{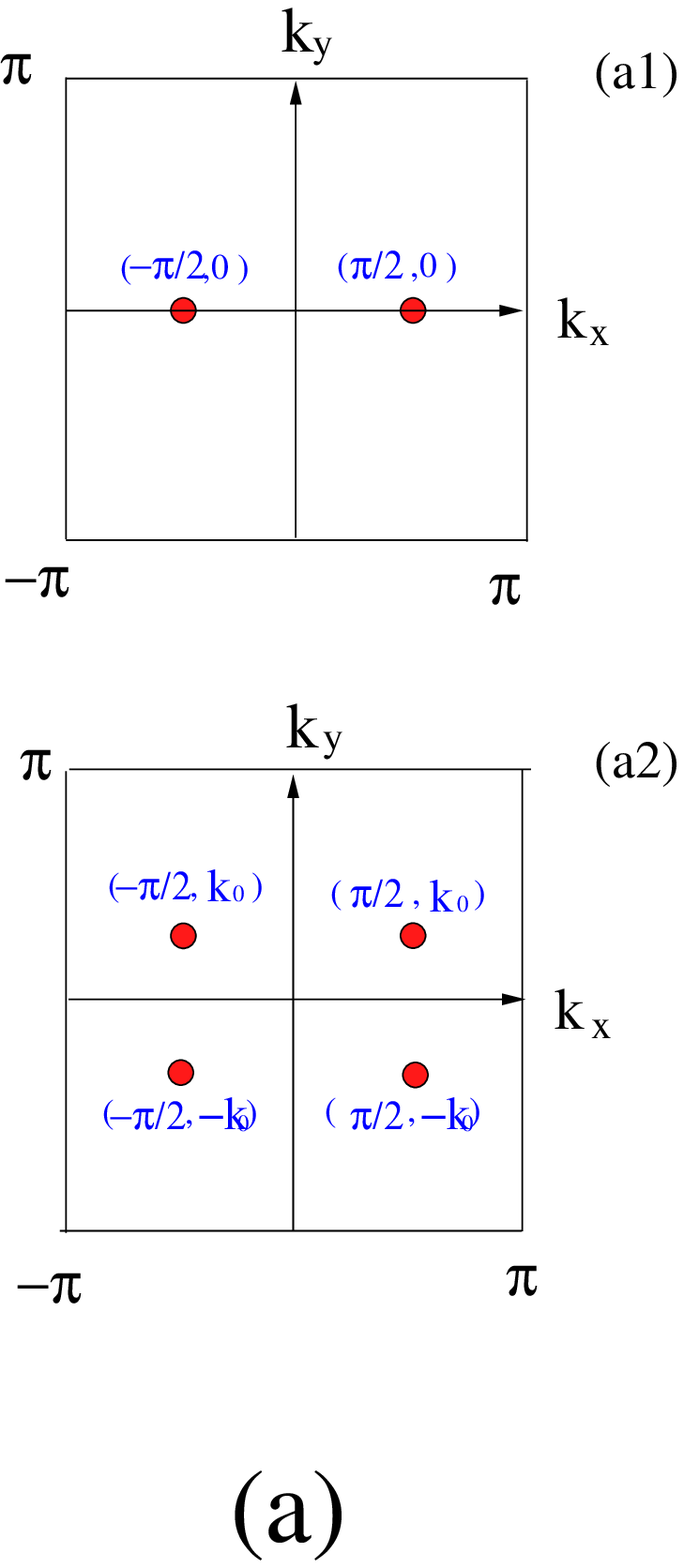}
\hspace{0.25cm}
\includegraphics[width=6cm]{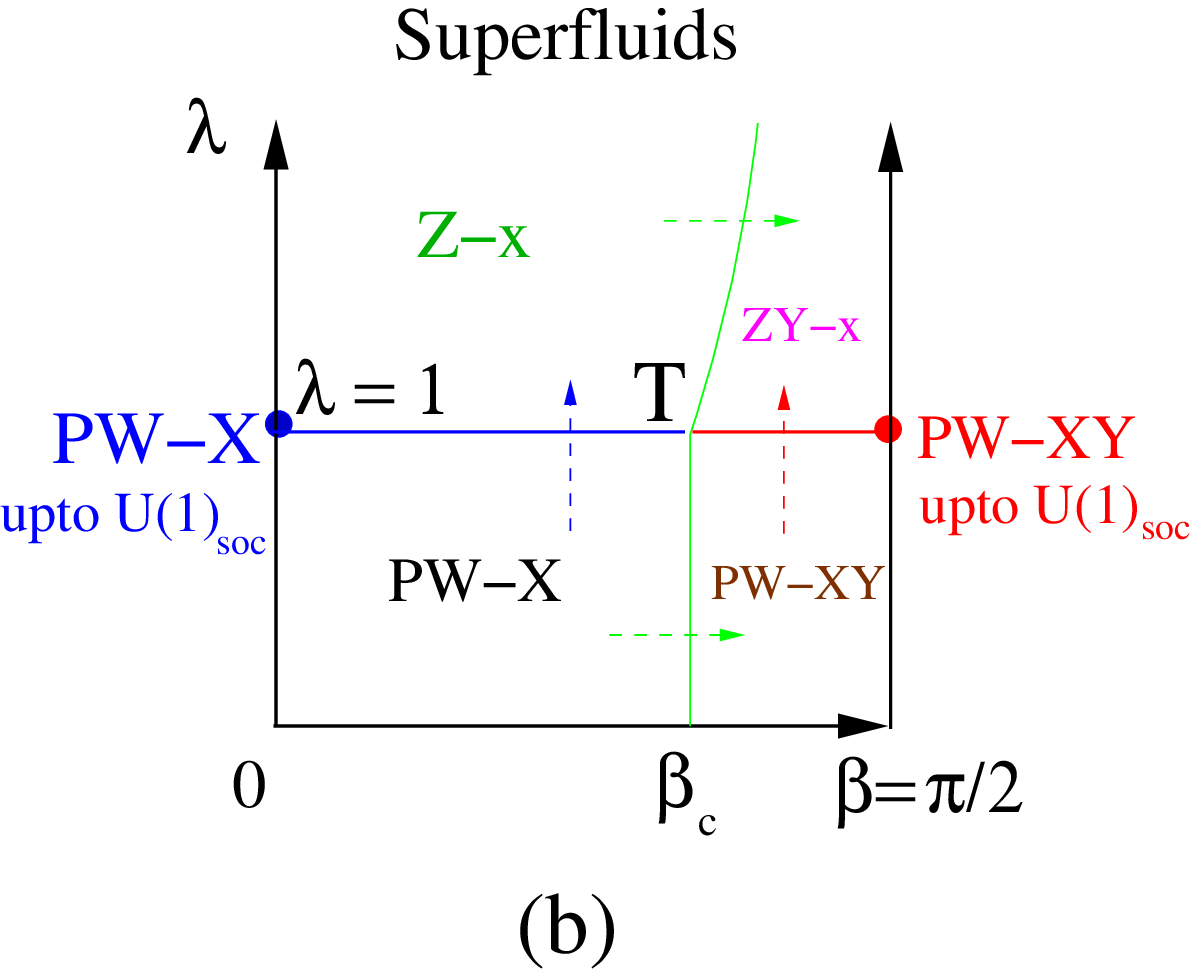}
\caption{ (a) Minima of kinetic energy along the anisotropic line
 $ ( \alpha=\pi/2, \beta ) $. All the minima are transferred to each other by the discrete symmetries of the Hamiltonian
(a1) two minima at $ (\pm \pi/2, 0) $ when $ 0 < \beta < \beta_c \sim 0.29 \pi $
(a2) four minima at $ (\pm \pi/2, k_0 ) $ when $ \beta_c  < \beta < \pi/2 $.
(b) The phase diagram in $ ( \beta, \lambda ) $ in the weak coupling limit $ U/t \ll 1 $. The system is always in a SF phase.
 There is a $ U(1)_{soc} $ symmetry at  $ \lambda= 1 $.
 This symmetry  is explicitly broken by the spin-dependent interaction  $ \lambda \neq 1 $.
 At $ \lambda < 1 $, there are  two plane wave (PW) SF phases: PW-X ( with  degeneracy $ d=2 $ )  and PW-XY ( $ d=4 $ )
 at $  0 < \beta < \beta_c $ and $ \beta_c  < \beta < \pi/2 $ respectively both of which break no translational symmetry.
 At $ \lambda > 1 $, there are two stripe SF phases along the x- bonds: Z-x  ( $ d=2 $ ) and
 ZY-x  ( $ d=4 $ ) at the two $ \beta $ intervals respectively. All the four phases meet at the Tetra-critical ( T ) point.
 At $ \lambda = 1 $, there are also two plane wave (PW) SFs:  PW-X and PW-XY both of which spontaneously
 breaks the $ U(1)_{soc} $ symmetry leading to one SOC Goldstone mode. All the other SFs at  $ \lambda = 1 $
 can be reached by performing a $ U(1)_{soc} $ transformation on the two representative PW SF states.
 The finite temperature phase transitions above all these SF phases are shown in Fig.\ref{rotondrive},\ref{sfdrive}. }
\label{sfall}
\end{figure}

 In this paper, we will investigate various spin-bond correlated Superfluids of  Rashba spin-orbit coupled (SOC) spinor bosons in a square lattice weakly interacting with a spin-anisotropic interaction parameter $ \lambda $.
 For the Rashba SOC, we focus on the anisotropic line $ (\alpha=\pi/2, \beta) $.
 We find the SOC provides a completely new frustrating source than the well studied geometric frustrations and Abelian flux.
 It leads to various novel frustrated Superfluids even in a bipartite lattice.
 Our main results are summarized in the $ ( \beta, \lambda ) $ phase diagram in the Fig.1.
 There are various SF states such as PW-X, Z-x, PW-XY and ZY-x phases \cite{notation}
 with not only off-diagonal long range order, but also different kinds of spin-bond correlated orders
 and various novel quantum  phase transitions among these SF phases.
 We develop a novel, general and systematic "order from quantum disorder"  analysis to  not only
 determine the quantum ground state, to calculate the roton gap, but also
 the correction to the spectrum in the PW-X state, especially the
 dispersion of the SOC Goldstone mode in the PW-X state at $ \lambda=1 $.
 When $ 0 < \beta < \beta_c $, by performing the analysis,
 we find the transition from the PW-X to the Z-x state is a novel second order one driven by the roton touchdown
 with different critical exponents on the two sides of the transition \cite{analogy}.
 The roton mode at $ \lambda < 1 $ becomes the SOC Goldstone mode at the
 spin isotropic interaction point $ \lambda=1 $ due to the $ U(1)_{soc} $ symmetry breaking.
 While the PW-X to the PW-XY is a second order Bosonic Lifshitz one with the anisotropic dynamic exponents
 $ (z_x=1, z_y=2) $ and an accompanying $ Z_2 $ transition breaking the $ {\cal P}_x $ symmetry in the spin sector. It is
 driven by the softening of the superfluid Goldstone mode along the $ y $ direction.
 The two second order transitions meet at a Tetra-critical point.
 We also extend our analysis to $  \beta_c  < \beta < \pi/2 $.
 The finite temperature phase transitions and the the transitions from weak to strong coupling
 above all these phases in Fig.\ref{sfall}b are sketched.
 The intimate analogy at $ \lambda=1 $  with the charge neutral Goldstone mode
 in the pseudo-spin sector in the Bilayer quantum Hall systems at the total filling factor $ \nu_T=1 $ are stressed.
 The transitions driven by the roton touchdown are contrasted with those leading to
 superfluid to supersolid transitions in frustrated lattices.
 Dramatic differences than the frustrated superfluids in the QAH systems of spinor bosons in a square lattice are listed.
 The experimental implications and detections of these novel phenomena in cold atoms are presented.

    We study short-range interacting bosons  hopping in a square lattice subject to  non-abelian gauge potential
\cite{rh}:
\begin{eqnarray}
H_{U} & = & -t\sum\limits_{\langle i,j\rangle}
b^\dagger(i\sigma)
U_{ij}^{\sigma\sigma'}
b(j\sigma')+h.c.    \nonumber  \\
 & + & \frac{U}{2} \sum_{i} ( n^2_{i \uparrow}+  n^2_{i \downarrow} + 2 \lambda  n_{i \uparrow} n_{i \downarrow} )
 - \mu \sum_{i} n_i
\label{intlambda}
\end{eqnarray}
    where the $ U_1=e^{i \alpha \sigma_x}, U_2=e^{i \beta \sigma_y} $ are the non-abelian gauge fields put on the two links in the square lattice ( Fig.1b).
    Setting $ \alpha=\beta=0 $, Eqn.\ref{intlambda} reduces to the conventional spin 1/2 ( two component ) spinor boson Hubbard model.
    In this paper, we focus on various novel Superfluids along the anisotropic SOC line \cite{rh} $ ( \alpha=\pi/2, \beta ) $
    and spin-anisotropic interaction $ \lambda $.

{\bf RESULTS }

{\bf 1. Mean field Analysis of the ground states in $ \beta, \lambda $ space. }
 In the weak interaction limit $ U \ll t $, one can diagonalize the kinetic energy first, then treat the $ U $ as a small perturbation.
 Along the anisotropic  line $ \alpha=\pi/2, 0 < \beta < \pi/2 $, the kinetic energy is:
\begin{equation}
	H_0=-t\sum_{i} [b_i^\dagger (i\sigma_x) b_{i+x}+b_i^\dagger e^{i\beta\sigma_y} b_{i+y}+h.c.]
\label{kinetic00}
\end{equation}
    which becomes  $ H_0=-2t\sum_{k}b_k^\dagger [\cos\beta\cos k_y-\sin k_x\sigma_x-\sin\beta\sin k_y\sigma_y] b_k $ in the
    $ \vec{k} $ space.

The non-interacting Hamiltonian $ H_0 $ has the following $ Z_2 $ reflection symmetry:
(1) $ {\cal P}_x $:  $ k_y \rightarrow - k_y, \sigma_y \rightarrow -\sigma_y,
\sigma_x \rightarrow \sigma_x,  \sigma_z \rightarrow -\sigma_z $.
(2) $ {\cal P}_y $ : $ k_x \rightarrow - k_x, \sigma_x \rightarrow -\sigma_x,
\sigma_y \rightarrow \sigma_y,  \sigma_z \rightarrow -\sigma_z $.
(3) $ {\cal P}_z $: $ k_x \rightarrow - k_x, \sigma_x \rightarrow -\sigma_x,
k_y \rightarrow - k_y, \sigma_y \rightarrow -\sigma_y, \sigma_z \rightarrow \sigma_z $.
This $ {\cal P}_z $ symmetry is also equivalent to a joint $ \pi $ rotation of orbital and spin around $ \hat{z} $ axis.
All these discrete symmetries are respected by the interaction.
In addition to the discrete symmetries, there is also a spin-orbital coupled $ U(1)_{soc} $ symmetry at $ \lambda=1 $.

When $0<\beta<\beta_c \sim 0.29 \pi $, there are two minima in the lower band at $(\pm\pi/2,0)$.
When $ \beta_c<\beta<\pi/2 $, there are four minima in the lower band at $(\pm\pi/2,\pm k_0)$ where $ \sin k_0=\sqrt{\sin^2\beta-\cot^2\beta}, 0<k_0<\pi/2$.
We will demonstrate our results on the left part $ 0 < \beta < \beta_c $ in Fig.\ref{sfall}.
Then at the end, we will outline our results to the right half  $ \beta_c < \beta < \pi/2 $.

    At the weak coupling, most of the bosons condense into:
\begin{align}
    \Psi_{i,0}=\frac{1}{\sqrt{2N_s}}
	\left[
	c_1e^{i\mathbf{K}\cdot\mathbf{r}_i}
	    \begin{pmatrix}
		1\\
		-1\\
	    \end{pmatrix}
	+c_2e^{-i\mathbf{K}\cdot\mathbf{r}_i}
	    \begin{pmatrix}
		1\\
		1\\
	    \end{pmatrix}
	\right]
\label{twonodes}
\end{align}
where $ \mathbf{K}=(\pi/2,0) $ and  $c_1$ and $c_2$ are two complex c-numbers
subject to the normalization condition $|c_1|^2+|c_2|^2=1$.

  For general $\lambda$, we obtain the mean field interaction energy of the state:
\begin{align}
    E^0_\text{int}
	=\frac{U}{2N_s}\left(1+\frac{\lambda-1}{2}[1-(c_1c_2^*+c_1^*c_2)^2]\right)
\label{E0}
\end{align}
  whose minimization leads to various mean field states.
In the following section, we study $\lambda<1$, $\lambda=1$, and $\lambda>1$ in detail.

{\bf 2. Order from disorder analysis on the PW-X SF at $  0 < \beta < \beta_c, \lambda<1$. }
The  minimization of Eq.\ref{E0} requires $c_1c_2^*+c_1^*c_2=0$ which implies two sets of solutions
(1) $ c_1=0 $  or $ c_2=0 $. (2) $ arg(c_1)-arg(c_2)=\pm \pi/2 $.
So one must perform an order from disorder analysis to find the quantum ground state
from this classically degenerate set of states. It turns out that due to the singularities at
the Plane wave solutions $ c_1=0 $  or $ c_2=0 $, it is difficult to perform such a analysis in the
original basis Eq.\ref{twonodes}. It is necessary to go to a different basis.
We find  it is convenient if we introduce two Z-x states which
satisfy $ Q \Psi^0_{\text{Z-x},\pm}= \pm \Psi^0_{\text{Z-x},\pm} $ with $ Q=\sum_i (-1)^{i_x} S^{z}_i $:
\begin{align}
    \Psi^{0}_{\text{Z-x},\pm}=
	\frac{1}{2\sqrt{N_s}}\left[
	    e^{i\mathbf{K}\cdot\mathbf{r}_i}
	    \begin{pmatrix}
		1\\
		-1\\
	    \end{pmatrix}
	    \pm
	    e^{-i\mathbf{K}\cdot\mathbf{r}_i}
	    \begin{pmatrix}
		1\\
		1\\
	    \end{pmatrix}
	\right]
\end{align}
    which can be used to re-parameterize Eq.\ref{twonodes} as:
\begin{align}
    \Psi_{i,0}	=
	e^{i\phi/2}\cos(\theta/2)\Psi^{0}_{\text{Z-x},+}
	+e^{-i\phi/2}\sin(\theta/2)\Psi^{0}_{\text{Z-x},-}
\label{p1}
\end{align}
which means that in Eq.\ref{twonodes}:
\begin{align}
    c_1  &=  [e^{i\phi/2}\cos(\theta/2)+e^{-i\phi/2}\sin(\theta/2)]/\sqrt{2},   \nonumber   \\
    c_2  &=  [e^{i\phi/2}\cos(\theta/2)-e^{-i\phi/2}\sin(\theta/2)]/\sqrt{2}
\label{c1c2}
\end{align}
  and the interaction energy becomes
\begin{align}
    E^{0}_\text{int}=\frac{U}{2N_s}\left(1+\frac{\lambda-1}{2}\sin^2\theta\right)
\label{classic}
\end{align}
  which determines $\theta=\pi/2,3\pi/2\cdots$. However, we cannot determine $\phi$ which forms a
  classically degenerate family. One must study quantum corrections to the ground state energy, namely,
  the order from disorder analysis.

Setting $\theta=\pi/2$, one can rewrite the spinor field into a condensation part plus a fluctuating part
$ \Psi=\sqrt{N_0} \Psi_0+ \psi $ where $N_0$ is the number of condensate atoms.
As shown in the Method section, the substitution
leads to the expansion of the Hamiltonian $ H=H^{(0)}+H^{(1)}+H^{(2)}+\cdots $
where the superscript denotes the order in the fluctuations $ \psi $ and $\cdots$ means high order.
$H^{(0)}=E^{(0)}$ is the classic ground state energy, setting $H^{(1)}=0 $ determines the value of the chemical potential
$ \mu=-2t(1+\cos\beta)+ Un_0(1+\lambda)/2 $. Diagonizing $ H^{(2)} $ by a $ 8 \times 8 $ Bogoliubov transformation leads to
\begin{align}
    H_\text{2}=E^{(2)}_{0}
	+\sum_{\mathbf{k}\in\text{RBZ}}
	 \sum_{l=1}^{4}\omega_l(\mathbf{k};\phi)
	 \left(\beta_{l,\mathbf{k}}^\dagger\beta_{l,\mathbf{k}}+\frac{1}{2}\right)
\label{h2}
\end{align}
  and the quantum corrections to the ground state energy:
\begin{align}
    E_\text{GS}(\phi)= E_{0t}
	+\frac{1}{2}\sum_{\mathbf{k}\in\text{RBZ}}\sum_{l=1}^{4}\omega_l(\mathbf{k};\phi)
\label{h2ground}
\end{align}
where $ E_{0t}=E^{(0)} + E^{(2)}_{0} $. We find that it picks up its minima at $\phi=0,\pi,\cdots$.
So the ``quantum order from disorder'' mechanism
selects the PW-X with $ c_1=1, c_2=0 $ or  $ c_1=0, c_2=1 $ among the classical degenerate family tuned by $ \phi $.

{\sl (a) The Excitation spectrum of the PW-X state }

  After determining the quantum ground state to be the PW-X state, we now investigate its excitation spectrum
  and also its critical behaviours when approaching the three boundaries $ \beta \rightarrow \beta^{-}_c $,
  $ \lambda \rightarrow 1^{-} $ and $ \beta \rightarrow 0^{+} $ in Fig.\ref{sfall}.

  There are two linear gapless modes located at\cite{subtract} $ (0,0) $ and $ (\pi,0 ) $ respectively:
\begin{eqnarray}
    \omega_G(\mathbf{q}) & = & \sqrt{n_0 t U(1+\lambda)[ q^2_x+ A(\beta)q_y^2] },   \nonumber  \\
     \omega_R(\mathbf{q}) & = & \sqrt{n_0 t U(1-\lambda)[ q^2_x+ A(\beta)q_y^2] }
\label{GR}
\end{eqnarray}
  where $ A( \beta )= \cos\beta-\sin^2\beta  $.
  The first is just the superfluid Goldstone mode, the second is the roton mode where  $ \mathbf{q}=\mathbf{k}-(\pi,0) $.
  As to be shown below, the roton mode will acquire a gap ( Fig.3a) from the order from disorder mechanism

  Because $ q_x $ direction is always non-critical, so we can set $ q_x=0 $ in the above equations.
  Because $ A( \beta_c )=0 $, so the Goldstone mode becomes quadratic along the $ q_y $ direction at $ \beta_c $:
\begin{equation}
   \omega_G(q_x=0, q_y) =\sqrt{n_0tU(1+\lambda) }/2 q^{2}_y,
\label{Gbetac}
\end{equation}
  which  is a bosonic Lifshitz transition with the anisotropic dynamic exponent $ z_x=1, z_y=2 $.

  As $ \lambda \to 1^{-} $, the Goldstone mode remains un-critical, but  the roton mode becomes a quadratic one:
\begin{equation}
   \omega^0_R(\mathbf{q})=t\sqrt{ [ q^2_x+ A(\beta)q_y^2] [  q_x^2+ A^{\prime}(\beta)q_y^2 ]}
\label{rotonq}
\end{equation}
  where $ A^{\prime}(\beta)=\cos\beta-\frac{\sin^2\beta}{1+n_0U/2t}  $.
  In the following, we will perform an order from  disorder analysis on the roton mode.

{\sl (b) The Roton mass gap generated from the "order by disorder" mechanism in the PW-X state }

\begin{figure}[!htb]
\centering
\includegraphics[width=0.225\textwidth]{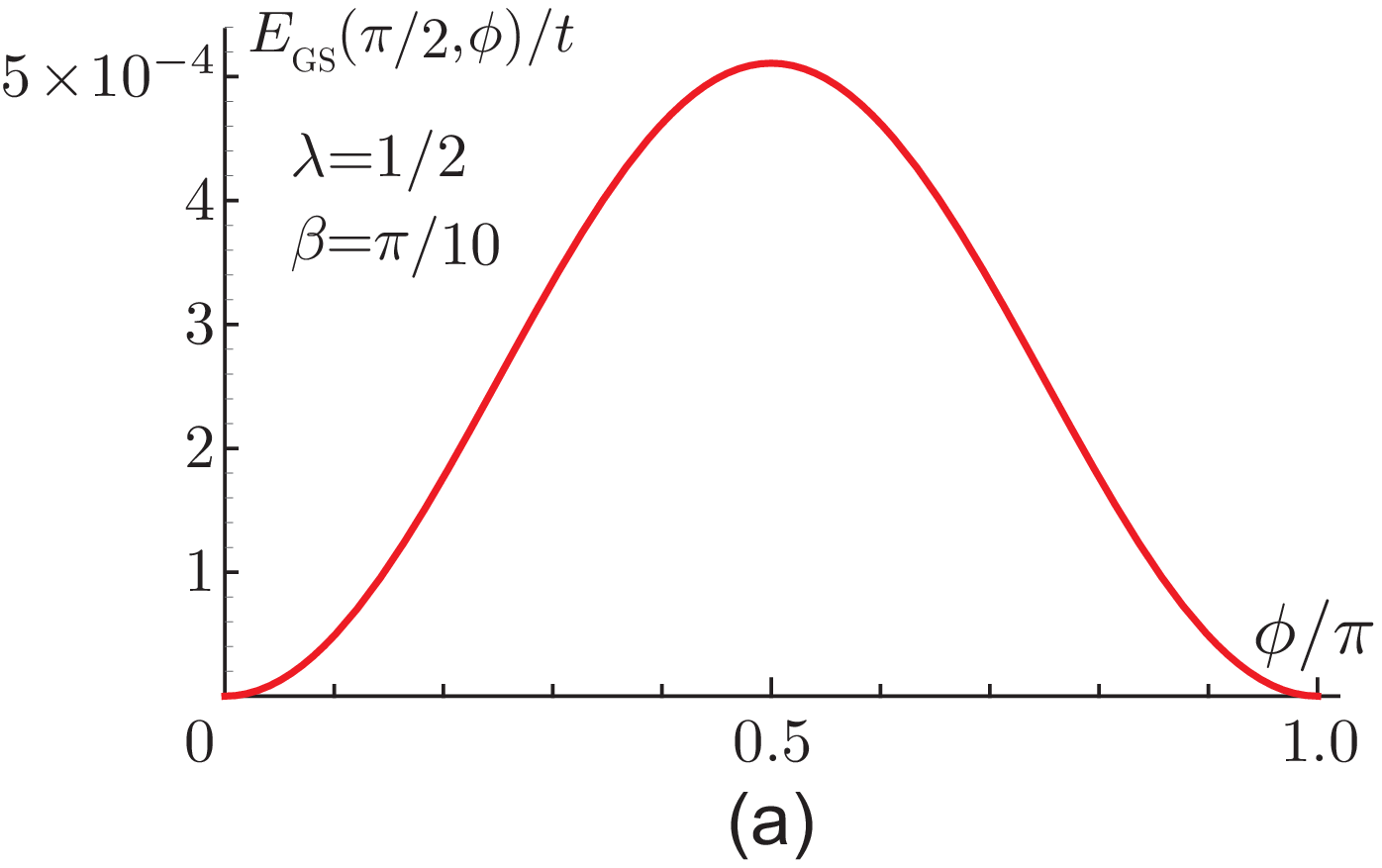}
\includegraphics[width=0.225\textwidth]{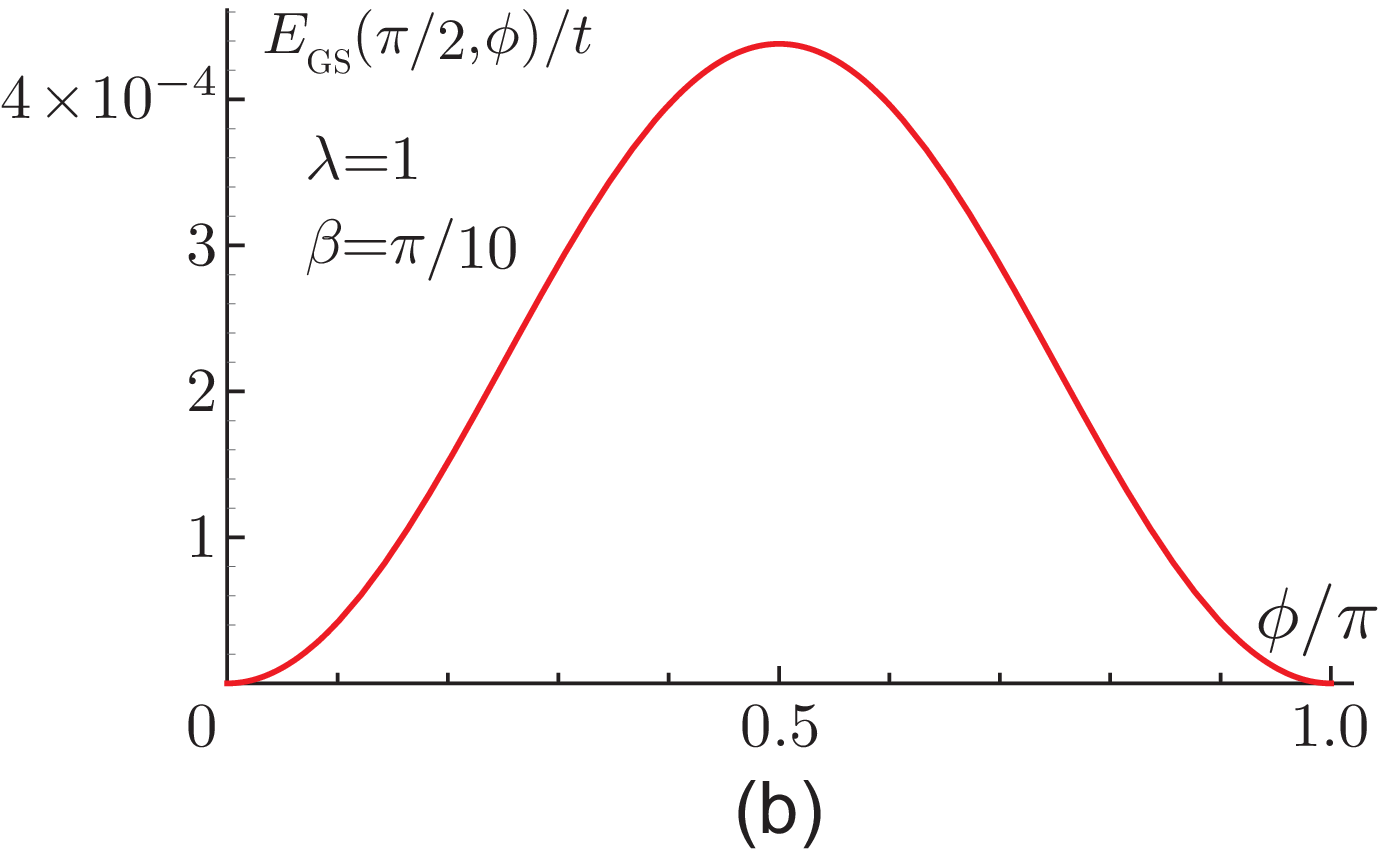}
    \caption{ The B term in the order from quantum disorder analysis (a) $ B ( \lambda < 1 ) $ defined in Eq.\ref{A2B2}.
    (b) $ B(\lambda = 1) $ defined in Eq.\ref{B2}.  }
\label{Bterm1}
\end{figure}

Combining Eq.\ref{classic} and Eq.\ref{h2ground},
we can extract the $\theta$ and $\phi$ dependence of the ground state energy
around its minimum:
\begin{align}
    H_2=E_{0t}+\frac{1}{2}A(\delta\theta)^2+\frac{1}{2}B(\delta\phi)^2
\label{A2B2}
\end{align}
where $\theta=\pi/2+\delta\theta$ and $\phi=0+\delta\phi$.
We obtain $A=(1-\lambda)Un_0^2/2$ analytically from Eq.\ref{classic},
and $B(\lambda) = b (n_0 U)^2/t \sim    $ can be numerically extracted from
Fig.\ref{Bterm1}a.

Because $ [ n_0\delta\theta/2,\delta\phi]= i \hbar $ forms a pair of conjugate variables.
The ``quantum order from disorder'' mechanism leads to the roton gap:
\begin{equation}
   \Delta^{-}_R=2\sqrt{AB}/n_0=\sqrt{2(1-\lambda)UB} \sim n_0U \sqrt{U/t}
\label{gap}
\end{equation}
which is shown in Fig.\ref{figgap}b to be a monotonically increasing function of $ \beta $.
When $ \beta \to 0^{+} $, the roton mode near $ (\pi,0) $ becomes the Goldstone mode due to the breaking of
$ \tilde{U}(1)_z $ at $ \beta=0 $.  Indeed,  the roton gap vanishes $ \beta \to 0^{+} $ as shown in Fig.4b.

As shown below, $B$ is non-critical as $\lambda$ approaches $ 1^{-} $,
so  critical behaviour of the roton gap $\Delta^{-}_R\propto\sqrt{1-\lambda}$ when $\lambda<1$.

In fact, as shown in the method section, the order from disorder mechanism not only leads to the
gap in Eq.\ref{gap}, but also modify the dispersion shown in Eq.\ref{gapdisp}

{\bf 3. The Order from disorder analysis on the PW-X SF at $ \beta_c < \beta < \pi/2, \lambda=1$.}
When $\lambda=1$ and $\beta<\beta_c$, Eq.\ref{E0} shows that any $c_1$ and $c_2$ give the same interaction energy,
thus they form a classically  degenerate ground-state manifold.
It was known that the Hamiltonian Eq.\ref{intlambda} at $\alpha=\pi/2$ and $\lambda=1$
has a spin-orbit coupled $U_\text{soc}(1)$ symmetry $ Q= \sum_i (-1)^{i_x} S^{y}_i $.
In order to distinguish between an accidental degeneracy and the exact degeneracy,
we introduce two Y-x states:
\begin{align}
    \Psi^0_{\text{Y-x},\pm}
    =\frac{1}{2\sqrt{N_s}}\left[
	e^{i\mathbf{K}\cdot\mathbf{r}_i}
	\begin{pmatrix}
	   1\\
	    -1\\
	\end{pmatrix}
	\pm
	ie^{-i\mathbf{K}\cdot\mathbf{r}_i}
	\begin{pmatrix}
	1\\
	1\\
	\end{pmatrix}
	\right]
\end{align}
 which satisfy $ Q \Psi^0_{\text{Y-x},\pm}= \pm \Psi^0_{\text{Y-x},\pm} $.

 They can be used to re-parameterize Eq.\ref{twonodes} as:
\begin{align}
    \Psi_{i,0}=
	e^{i\phi/2}\cos(\theta/2)\Psi^{0}_{\text{Y-x},+}
	+e^{-i\phi/2}\sin(\theta/2)\Psi^{0}_{\text{Y-x},-}
\label{p2}
\end{align}
which means that in Eq.\ref{twonodes}:
\begin{align}
    c_1=[(e^{i\phi/2}\cos(\theta/2)+e^{-i\phi/2}\sin(\theta/2)]/\sqrt{2},    \nonumber  \\
    c_2=i[(e^{i\phi/2}\cos(\theta/2)-e^{-i\phi/2}\sin(\theta/2)]/\sqrt{2}
\end{align}

The $ U(1)_{soc} $ symmetry at $ \lambda=1 $ dictates that the energy should be independent of $ \phi $.
We expect that the classical degeneracy at the angle $ \theta $ will be lifted by quantum effects,
and a unique $\theta$ will be determined by the order-from-disorder mechanism.

 We set $ \phi=0 $ to spontaneously breaks the $ U(1)_{soc} $ symmetry.
 Following the similar procedures as in the last section, we obtain Eqs.\ref{h2},\ref{h2ground}
 which pick up its minima at $\theta=\pi/2,\cdots$.
 So the ``quantum order from disorder'' mechanism still selects the PW-X state
 $(\theta=\pi/2,\phi=0)$. All the other exactly degenerate states can be generated by changing $ \phi $.
 If we take $ c_1=1, c_2=0 $ PW-X state in Eq.\ref{twonodes}, then the $ U(1)_{soc} $ related family is
\begin{align}
    \Psi^0 ( \phi )
    =\frac{1}{\sqrt{2N_s}}\left[ \cos \phi
	e^{i\mathbf{K}\cdot\mathbf{r}_i}
	\begin{pmatrix}
	   1\\
	    -1\\
	\end{pmatrix}
	-\sin \phi
	e^{-i\mathbf{K}\cdot\mathbf{r}_i}
	\begin{pmatrix}
	1\\
	1\\
	\end{pmatrix}
	\right]
\label{socu1}
\end{align}
  whose corresponding spin-orbital structure is $  \langle \sigma_x \rangle = -\cos 2\phi,
  \langle \sigma_y \rangle = 0, \langle \sigma_z \rangle = -(-1)^x \sin 2 \phi $.
  Setting $ \phi=0, \pi/2 $  and $ \phi=\pi/4, 3 \pi/4 $ recovers the two PW-X state  and the Z-x,${\pm}$ state respectively.

By comparing $\Psi_i^{\lambda<1}(\theta_1=\pi/2,\phi_1)$ in Eq.\ref{p1}
with $\Psi_i^{\lambda=1}(\theta_2,\phi_2=0)$ in  Eq.\ref{p2}
( we used the subscript $1$ for $\lambda<1$ and $2$ for $\lambda=1$ case ), we can see the relations
between the order from disorder variables in the two cases:
\begin{equation}
    \theta_2=\pi/2-\phi_1
\label{exch}
\end{equation}
 which just shows the $ ( \theta, \phi) $ exchanges in the Z-x basis in Eq.\ref{p1} and the $ Y-x $ basis in \ref{p2}.

{\sl (a) The Goldstone mode spectrum computed from the "order by disorder" mechanism in the PW-X SF state }

 The $U_\text{soc}(1)$ symmetry dictates no $\phi$ dependence, so the ground state energy an be written as:
\begin{align}
    E_{GS}( \theta ) =E_{1}+\frac{1}{2}B(\delta\theta)^2
\label{B2}
\end{align}
where $ E_{1}= E_{GS}( \theta=\pi/2 ) $ and $ \delta\theta =\theta-\pi/2 $ and  $ B(\lambda=1)$ can be extracted
from Fig.\ref{Bterm1}b.
 The $A$ term in Eq.\ref{A2B2}  vanishes due to the exact $U_\text{soc}(1)$ symmetry.
 Because $\lim_{\lambda\to1} B(\lambda<1)=B(\lambda=1)$, so $B(\lambda)$ is continuous at $\lambda=1$.

 As shown in the method section, the ``quantum order from disorder'' mechanism
changes the quadratic dispersion in Eq.\ref{rotonq} to  a linear one:
which is nothing but the SOC Goldstone mode due to the spontaneous breaking of the $U_\text{soc}(1)$ symmetry,
\begin{align}
    \omega_R=\sqrt{\frac{2Bt}{n_0}\left[q_x^2 +  A^{\prime}(\beta)  q_y^2\right]}
\label{rotonl}
\end{align}
   where because $ B \sim (n_0U)^2/t $, so the slope of the roton mode $ v \sim \sqrt{n_0} U $ is smaller than the
   superfluid velocity $ c \sim \sqrt{n_0 U t } $ in Eq.\ref{GR} ㄗ Fig.3bㄘ.
   Due to the wide momentum separation between the Goldstone and roton mode, the Goldstone mode is
   not affected by the order from disorder analysis.

   We observe that at the bosonic Lifshitz transition point $ \beta_c $ in Fig.1bㄛ $ A(\beta_c)=0 $, but $ A^{\prime}(\beta_c)> 0 $,
   so the SF Goldtone mode in Eq.\ref{GR} becomes quadratic signifying the transition,
   while the SOC Goldstone mode in Eq.\ref{rotonl} remains linear along the $ y $ direction.
   So the SOC Goldstone mode is non-critical across the  bosonic Lifshitz transition.
   This fact is important to determine the finite temperature phase transitions above the $ \lambda=1 $ line in Fig.7.


\begin{figure}[!htb]
\centering
    \includegraphics[width=0.45\textwidth]{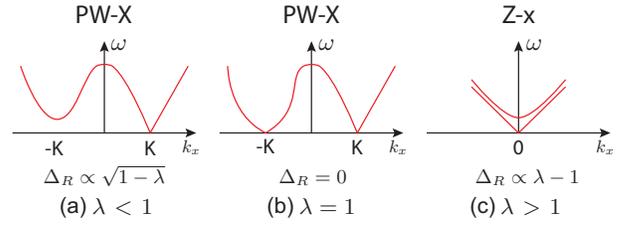}
    \caption{Evolution of the low energy excitations as $\lambda$ changes from $ \lambda < 1 $ to $ \lambda = 1 $ and $ \lambda > 1 $
    at a fixed $ 0 < \beta < \beta_c $.
    (a) The roton gap $ \Delta^{-}_R \sim \sqrt{1-\lambda} $ generated by the order from disorder mechanism at $ \lambda < 1 $.
    (b) The SOC Goldstone mode at $ (-\pi/2,0) $ due to the $ U(1)_{soc} $ breaking computed
    by the order from disorder mechanism at $ \lambda = 1 $.
    (c) The  roton gap $ \Delta^{+}_R \sim \lambda-1 $ in the Z-x state at $ \lambda >1 $.
     The superfluid Goldstone mode at $ (\pi/2,0) $ from the $ U(1)_c $ breaking remains uncritical through the transition. }
\label{twoexp}
\end{figure}

{\bf 4. The Z-x SF state at $ \beta_c < \beta < \pi/2, \lambda > 1$. }
When $\lambda>1$ and $\beta<\beta_c$, the  minimization of Eq.\ref{E0} requires $c_1c_2^*+c_1^*c_2=\pm1$
which means $c_1=\pm c_2=e^{i\gamma}/\sqrt{2}$ where $\gamma$ is a globe phase.
So  the 2 fold degenerate  ground state is  either $\Psi_{\text{Z-x},+}$ with $c_1=c_2=1/\sqrt{2}$  or $\Psi_{\text{Z-x},-}$ with $c_1=-c_2=1/\sqrt{2}$.

Let us pick up the $\Psi_{\text{Z-x},+}$ state, we find its 4 excitation modes $\omega_{1,2,3,4}(\mathbf{k})$ in ascending order:
$\omega_1(\mathbf{k})$ contains a Goldstone mode at $(0,0)$ ( Fig.3c),
$\omega_2(\mathbf{k})$ contains a roton mode at $(0,0)$ ( Fig.3c)
and $\omega_{3,4}(\mathbf{k})$ are high energy bands ( not shown in Fig.3c ).
 Note that the Z-x state breaks the translation symmetry to two sites per unit cell, so
 the Brillioun zone ( BZ) in the PW-X state  becomes the Reduced BZ (RBZ)
 with $ -\pi/2 < k_x < \pi/2, -\pi < k_y < \pi $.

We also extract the long wavelength limit of the SF Goldstone mode:
\begin{align}
    \omega_1(\mathbf{k})
	=\sqrt{2n_0t U\left[
		k_x^2+ A^{\prime \prime}(\beta,\lambda) k_y^2		\right]}
\end{align}
  where $ \mathbf{k} \in RBZ $ and $ A^{\prime \prime}(\beta,\lambda) = \cos\beta-\frac{\sin^2\beta}{1+Un_0(\lambda-1)/4t} $.
  Because $ A^{\prime \prime}(\beta,\lambda=1) = A( \beta) $, it recovers the SFGoldstone mode in Eq.\ref{GR} at $ \lambda=1 $.
  However, when $ \lambda >1 $, by setting $ A^{\prime \prime}(\beta_0,\lambda)=0 $,
  one can see the Bosonic Lifshitz transition line shifts to
  $ \beta_0 ( \lambda ) > \beta_c $ as shown in Fig.1b.
  We also obtain the critical behaviour of the roton mode $\Delta^{+}_R=\omega_2(0,0)\propto\lambda-1$.

\begin{figure}[!htb]
\centering
\includegraphics[width=\linewidth]{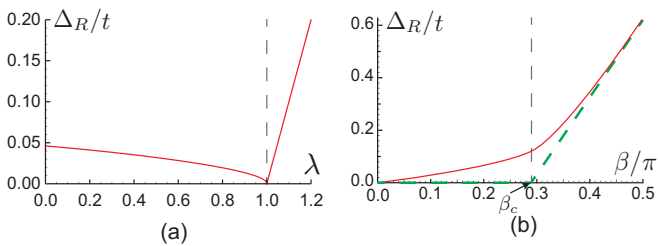}	
    \caption{(a) The quantum critical behaviour of the roton gap $\Delta_R \sim |1-\lambda|^{\beta^{\prime}} $
     as a function of $\lambda$ at a fixed $\beta=\pi/10$. $ \beta^{\prime}_-=1/2, \beta^{\prime}_+=1 $ provide the first example
     of a QCP with different critical exponents on two sides of a QCP.
     However, as shown in Sec.4, the transition from the PW-XY to the ZY-x is also driven by the roton touchdown with
     equal exponents on both sides $ \beta^{\prime}_- = \beta^{\prime}_+=1 $.
	(b) The roton gap $\Delta_R$ in red line calculated by order from disorder analysis Eq.\ref{gap} in the PW-X and PW-XY
        state when $ \lambda < 1 $ is a monotonically increasing function of $\beta$ at a fixed $\lambda=1/2$.
        The Green line is before the order from disorder analysis.
	    Both fixed $t=1$, $n_0U=1$ and $n_0\approx 1$. Note that $\beta_c\approx0.288\pi$. }
\label{figgap}
\end{figure}


{\bf 5. The SFs and transitions on  $  \beta_c < \beta < \pi/2 $. }

So far, we focused on the left part of the phase diagram Fig.1.
In this section, we apply similar methods to study the SFs and transitions on the right half  $    \beta_c < \beta < \pi/2 $ of Fig.\ref{sfall}.

{\sl (a)  When $  \lambda < 1  $ }, the $ \vec{Q} $ takes one of the four minima $ (\pm \pi/2, \pm k_0 ) $ shown in Fig.1a2
  and the $ \phi(\beta) $ is determined by $ k_0 $﹛and given in Eq.\ref{spinor0}.
  Then  $ \langle \sigma_x \rangle = -\cos \phi(\beta), \langle \sigma_y \rangle = \sin \phi (\beta),
  \langle \sigma_z \rangle = 0 $. This is the PW-XY state in Fig.\ref{sfall}b where the spin is in the $ XY $ plane.
  It breaks the  $ {\cal P}_x $ symmetry.
  There is a quantum phase transition from the plane wave superfluid at $ \beta < \beta_c $ ( PW-X) to the plane wave superfluid at
  $  \beta_c < \beta < \pi/2 $ (PW-XY) which is in the Ising university class in the spin sector.
  As shown in Sec.2 and confirmed also below, this Iisng transition in the spin sector is just affiliated to
  the bosonic Lifshitz transition at $ \beta=\beta_c $ with the dynamic exponents $ ( z_x=1, z_y=2 ) $.

Writing $ \Psi=\Phi_{0,PW-XY} + \psi $, following similar procedures as in the PW-X phase,
we are able to find the excitation spectrum above the PW-XY state.
We find  a gapless SF mode \cite{subtract} at $ (0,0) $ which is due to the $ U(1)_c $ symmetry breaking.
We also find three gapped modes at $ (\pi,0 ) $, $ (\pi, 2 k_0) $ and $ (0, 2k_0) $ ( Fig.1a2).
In the following, we will study the critical behaviors of these modes (1) across the phase transition from the PW-XY to
the ZY-x transition through the $ \lambda =1 $ line where there is the enhanced $ U(1)_{soc} $ symmetry.
(2) across the phase transition from the PW-X to the PW-XY  through the bosonic Lifshitz transition at $ \beta=\beta_c $.
(3) approaching to the right axis $ \beta \rightarrow \pi/2^{-} $.

At a fixed  $ \beta_c < \beta < \pi/2 $, as $ \lambda \to 1^{-} $ along the red dashed line in Fig.1b,
the gapless linear SF mode remain un-critical,
the roton  gap at $ (\pi,0 ) $ decreases to zero as $ \Delta^{-}_{R} \sim 1- \lambda $, the mode
 becomes linear, it is just the SOC Goldstone mode due to the $ U(1)_{soc} $ symmetry breaking at $ \lambda=1 $.
We find the roton gap at $ (\pi, 2 k_0 ) $ also decreases to zero, but the mode becomes quadratic.
As to be shown below, an order from disorder analysis at $ \lambda=1 $ will open a gap to this quadratic mode.
However, the roton gap at $ (0,2 k_0 ) $ remains  as $ \lambda \to 1^{-} $.

At a fixed  $ \lambda<1  $, as $ \beta \to \beta^{+}_c $ in Fig.1b,
the gapless linear SF mode at $ (0,0 ) $ also becomes quadratic along the $ y $ direction as in Eq.\ref{Gbetac}, so confirm
the transition at $ \beta_c $ is a second order bosonic Lifshitz one with the dynamic exponents $ ( z_x=1, z_y=2 ) $.
The roton  gap at $ (\pi,0 ) $ decreases to zero, the mode becomes linear, matches the behaviours of the roton mode
from the left ( before applying the order from disorder analysis ) shown in Fig.\ref{figgap}b.
Of course, as  $ \beta \to \beta^{+}_c $, then $ k_0 \to 0 $,
the modes at  $ (\pi, 2 k_0 ) $ and $ (0,2 k_0 ) $ become the same as those at $ (\pi, 0) $ and $ (0, 0 ) $ respectively.
As $ \beta \to \pi/2^{-} $ in Fig.1b,   $ k_0 \to \pi/2 $, only the mode at $ (\pi,\pi) $ becomes a gapless linear one
( as to be shown in the next section, this mode becomes the Goldstone mode due to the $ \tilde{\tilde{U}}_z(1) $ symmetry breaking
  at $ \lambda < 1, \beta =\pi/2 $ ), while the modes at $ (0,\pi) $ and $ (\pi, 0 ) $ remain gapped.

 {\sl (b) Order from quantum disorder at $  \lambda = 1  $. }
   We find that in addition to the exact  $ U(1)_{soc} $ symmetry,
   there is also a spurious $ U(1) $ symmetry leading to a classically degenerate family of states.
   By a similar order from disorder analysis as in the $  \beta < \beta_c, \lambda=1 $ case, we
   show that the quantum ground state is the PW-XY state upto all the other states related
   by the exact $ U(1)_{soc} $ symmetry  ( Fig.1b ).  Then there exists two gapless modes which are
   due to the $ U(1)_c $ and $ U(1)_{soc} $ symmetry breaking respectively.

   Indeed, as approaching from below  $ \lambda \to 1^{-} $, as said above, the roton gap $ \Delta_R $
   at $ (\pi,0 ) $ decreases to zero as $ \Delta^{-}_R \sim 1- \lambda $,
   the mode becomes linear at $ \lambda \to 1 $ which is precisely the SOC Goldstone mode due to the $ U(1)_{soc} $ symmetry breaking.
   The roton gap at $ (\pi, 2 k_0 ) $ also decreases to zero, but the mode becomes quadratic which
   is due to the spurious $ U(1) $ symmetry at $ \lambda=1 $.
   The order from disorder analysis at $ \lambda=1 $ opens a gap to the quadratic mode at $ (\pi, 2 k_0 ) $.
   So $ (\pi, 2 k_0 ) $ remains un-critical across the PW-XY to the ZY-x transition.

   As  $ \lambda \to 1^{-}, \beta \to \pi/2  $, namely approaching the Abelian point with a $ \pi $ flux,
   all the three modes touch zero: the modes at $ (0, \pi), ( \pi,0) $ become linear, while
   the $ ( \pi,\pi) $ mode becomes quadratic ( which, as to be shown below,
   will become a linear mode after an order from disorder analysis ).
   These facts recover those found at the $ \pi $ flux Abelian point in \cite{abesf}
   and will be discussed again in Sec.6.


 {\sl (c)  When $  \lambda > 1  $ }.  $ \vec{Q}_1 $ can take one of the two values $ \vec{Q}_1=( \pi/2, \pm k_0 ) $ in Fig.\ref{sfall}a2,
  then  $ \vec{Q}_2=( -\pi/2, \pm k_0 ) $.
  It leads to the 4-fold degenerate ZY-x phase with the momentum $ (\pi,0 ) $ in Fig.\ref{sfall}b where the spin is
  in the $ ZY $ plane as shown in Eq.\ref{ZYx} and Fig.8.
  Following the similar procedures as in the Z-x state on the left in Fig.1b, in the RBZ, we find a SF Goldstone mode
  $ \omega_1(\mathbf{k}) $ and also a roton mode $ \omega_2(\mathbf{k}) $ with a roton gap $\Delta^{+}_R=\omega_2(0,0)\propto\lambda-1$.

{\bf 6. The physical picture from the left and right Abelian points }

{\sl (a) Left Abelian point: The analogy  at $ \lambda=1 $ to the neutral gapless mode in the Bilayer Quantum Hall system (BLQH) }

 At the left Abelian point $ (\beta=0, \lambda=1 ) $, the spin $ SU(2) $ symmetry becomes evident in the rotated basis $\tilde{\mathbf{S}}_i=( S_i^x,(-1)^{i_x}S_i^y, (-1)^{i_x} S_i^z)$. The breaking of this $ \tilde{SU}(2) $
 leads to the quadratic FM spin wave mode $ \omega \sim k^2 $. However,  at any $ \beta >0 $,
 the spin $ \tilde{SU}_s(2) $ symmetry is reduced to
 the $ U(1)_{soc} $. So the FM spin wave becomes a linear Goldstone mode at any $ \beta >0 $.
 Very similar phenomena happen in the bilayer quantum Hall systems (BLQH) at the total filling factor $ \nu_T=1 $\cite{blqhrev,blqhye}.
 At $ d=0 $, there exists an exact $ SU_s(2) $ symmetry dictating  a $ \omega \sim k^2 $ neutral gapless mode.
 At any finite distance $ d > 0 $, a capacitive term reduces the spin $ SU(2) $ symmetry to a easy plane $ U(1) $ symmetry, therefore
 transfers the quadratic FM mode to a linear Goldstone mode due to the breaking of the $ U(1) $ symmetry.
 In BLQH, the capacitive term is due to the difference between inter- and intra-layer Coulomb interaction
 which explicitly breaks the $ SU(2) $ to $ U(1) $ at any $ d >0 $.
 Here the capacitive term Eq.\ref{B2} is generated by the order from disorder mechanism.
 As shown in Sec.3 and detailed in the SM, it is this capacitive term which transforms the otherwise quadratic dispersion
 to the linear Goldstone mode as dictated by the above exact symmetry breaking analysis.

 Of course, in the $\tilde{\mathbf{S}} $ basis, moving up along the left axis ( $ \lambda > 1 $ ) is the Ising limit,
 down ( $ \lambda < 1 $ ) is the easy plane limit where the $\tilde{U}(1)_z $ symmetry breaking leads to the spin Goldstone mode.
 Using the relation $ \tilde{S}_i^z = (-1)^{i_x} S_i^z $, this Goldstone mode becomes the one at $ (\pi,0) $
 in the original basis, which is precisely observed in the PW-X phase in the limit $ \lambda < 1, \beta \to 0^{+} $ as discussed in Sec.2
 Indeed,  the roton gap vanishes $ \beta \to 0^{+} $ as shown in Fig.\ref{figgap}a.

{\sl (a) Right  Abelian point:  Four linear gapless modes reduce to two at $ \lambda=1 $.  }

 In the right $ \pi $ flux Abelian point $ (\beta=\pi/2, \lambda=1 ) $, the spin $ SU(2) $ symmetry becomes evident
 in the rotated basis  $\tilde{\tilde{\mathbf{S}}}_i=((-1)^{i_y}S_i^x,(-1)^{i_x}S_i^y,(-1)^{i_x+i_y}S_i^z)$
 which was used in \cite{rh,notesy}. In the non-Abelian gauge  at $ (\alpha=\pi/2, \beta=\pi/2 ) $,
 we find \cite{junsen} a spurious $ U(1) $ symmetry which leads to a degenerate family of states
 connecting $ \mathbf{K}_1=(\pi/2,\pi/2) $ and $ \mathbf{K}_3 =- \mathbf{K}_1 $.
 A similar order from disorder analysis as in Sec.2
 shows that quantum corrections select the PW-XY state to be the ground state upto the spin $ SU(2) $ symmetry.
 Taking the the PW-XY state as the simplest quantum ground state, we find
 there are three gapless linear modes at $ \mathbf{K}_1=(\pi/2,\pi/2) $, $ \mathbf{K}_2 $ and $ \mathbf{K}_4 $,
 one quadratic mode at $ \mathbf{K}_3  $.
 By performing a similar order from quantum disorder analysis \cite{junsen} as that in Sec.3, we find the quantum corrections
 change the quadratic mode at $ \mathbf{K}_3 $  to a linear one. This is consistent with the results achieved
 in  \cite{abesf} using a Abelain gauge which explicitly keeps the spin $ SU(2) $ symmetry.
 In \cite{abesf}, we showed that there are strong quantum fluctuations induced by the coupling between the off-diagonal long range order
 in the SF sector and the magnetic orders in the spin order ( This is in sharp contrast to
 the case at the left Abelian point discussed above where there is no such coupling.)
 The complete symmetry breaking $ U(1)_c \times SU(2)_s \rightarrow 1 $ leads to
 4 linear gapless modes. When transforming the $ \theta=\pi $ Abelian gauge to the present
 non-abelian gauge with $ ( \alpha=\pi/2, \beta=\pi/2 ) $,
 the $ 90^{\circ} $ co-planar state in \cite{abesf} becomes the PW-XY state,
 the 4 linear modes are shifted to the 4 minima at $ (\pm \pi/2, \pm \pi/2) $.

 When slightly away from the right $ \pi $ flux Abelian point, the 4 minima shift to $ (\pm \pi/2, \pm k_0 ) $,
 only 2 of the 4 gapless modes survive, one due to the $ U(1)_c $ symmetry breaking at (0,0), another due to the
 $ U(1)_{soc} $ symmetry breaking at $ (\pi,0) $. While the modes at $ (0, 2 k_0 ) $ and $ (\pi, 2 k_0 ) $ acquire gaps.

 Of course, in the $ \tilde{\tilde{\mathbf{S}}} $ basis, moving up along the right axis ( $ \lambda > 1 $ ) is the Ising limit,
 down ( $ \lambda < 1 $ ) is the easy plane limit where there is another spin Goldstone mode due to the $\tilde{\tilde{\mathbf{U}}}(1)_z $ symmetry breaking.
 Using the relation $ \tilde{\tilde{S}}_i^z = (-1)^{i_x+i_y } S_i^z $, this spin Goldstone mode becomes the one at $ (\pi,\pi) $
 in the original basis, which is precisely observed in the PW-XY phase in the limit
  $ \lambda < 1, \beta \to \pi/2^{-} $ as discussed in Sec.5.
 Indeed, in this limit, only $ ( 0,0 ) $ and $ (\pi, \pi ) $ modes become linear gapless ones,
 but $ (\pi,0) $ and $ ( 0,\pi) $ remain gapped.

\begin{figure}[!htb]
\centering
    \includegraphics[width=0.225\textwidth]{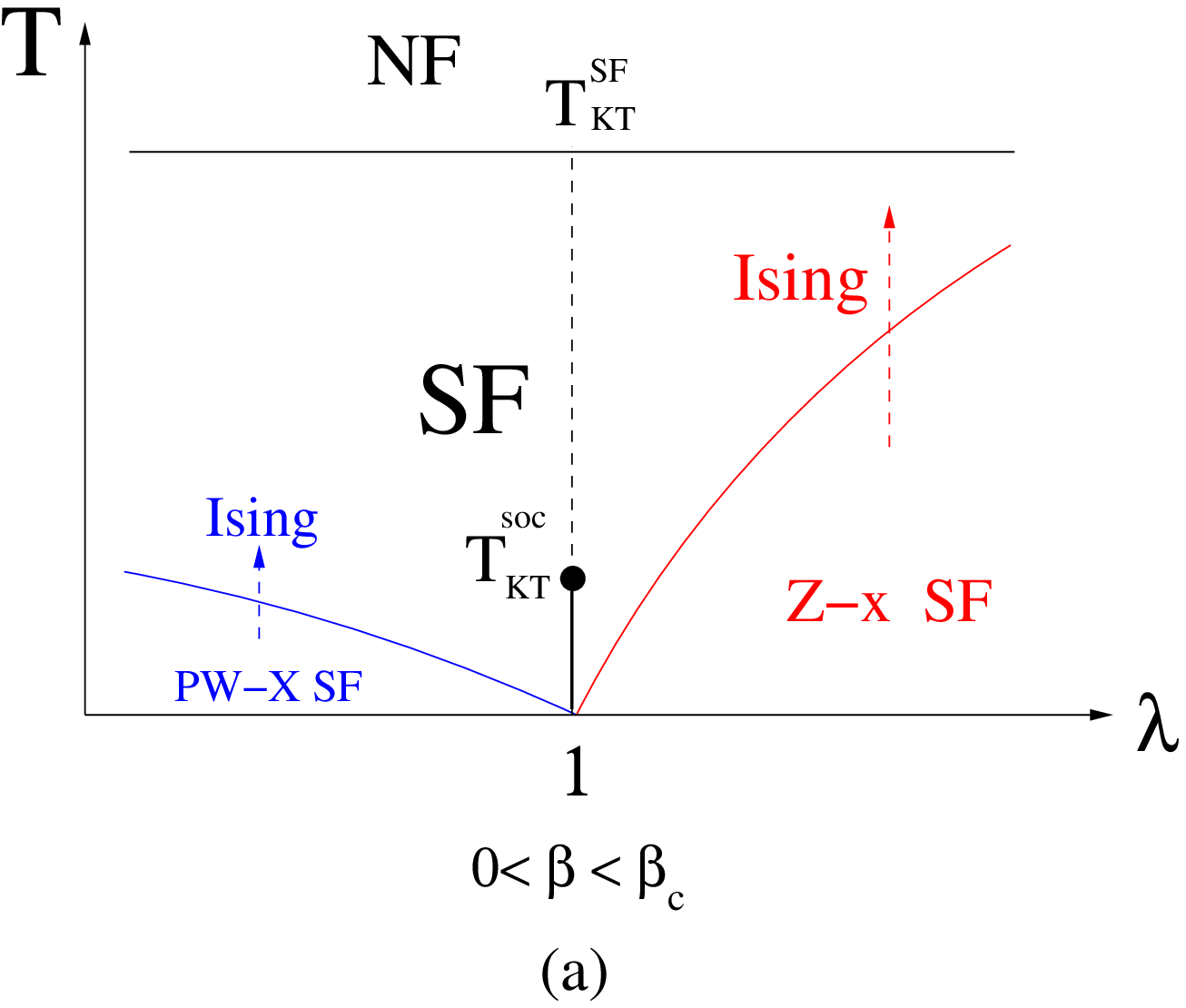}
    \includegraphics[width=0.225\textwidth]{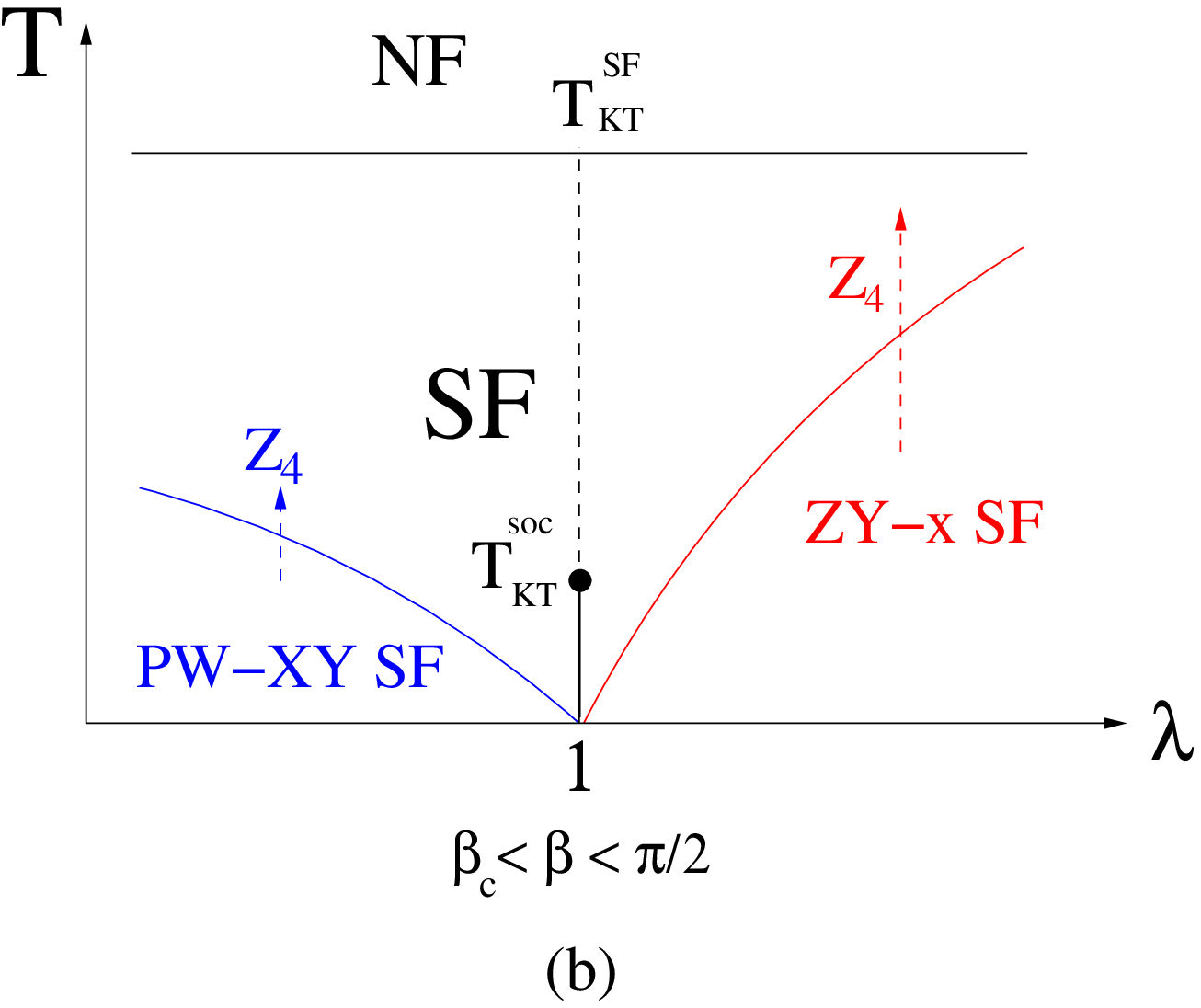}
    \caption{ The quantum phase transitions at $ T=0 $ are driven by roton touchdowns.
     The finite temperature phase transitions in
    (a) $ 0 < \beta < \beta_c $. There is a Ising melting transition in the PW-X SF and Z-x SF
      respectively.  The transition temperature $ T_2 \sim \Delta_R \sim |1- \lambda |^{\beta^{\prime} } $ scales as
      the roton gap at $ (\pi,0) $ in Fig.\ref{figgap}a.
      There is a KT transition at $ T^{soc}_{KT} $ to restore the $ U(1)_{soc} $ symmetry at $ \lambda=1 $.
      Then there is another higher KT transition at $ T^{SF}_{KT} $ from the SF to a normal fluid (NF).
    (b) $ \beta_c < \beta < \pi/2 $, There is  a $ Z_4 $ clock melting transition in the PW-XY SF and ZY-x SF
      respectively. The transition temperature $ T_4 \sim \Delta_R \sim |1- \lambda | $ scales as  the roton gap at $ (\pi,0) $.  }
\label{rotondrive}
\end{figure}

{\bf 7. Finite temperature phase transitions above all the SF phases and transitions from the weak
to strong coupling }

\begin{figure}[!htb]
\centering
    \includegraphics[width=0.225\textwidth]{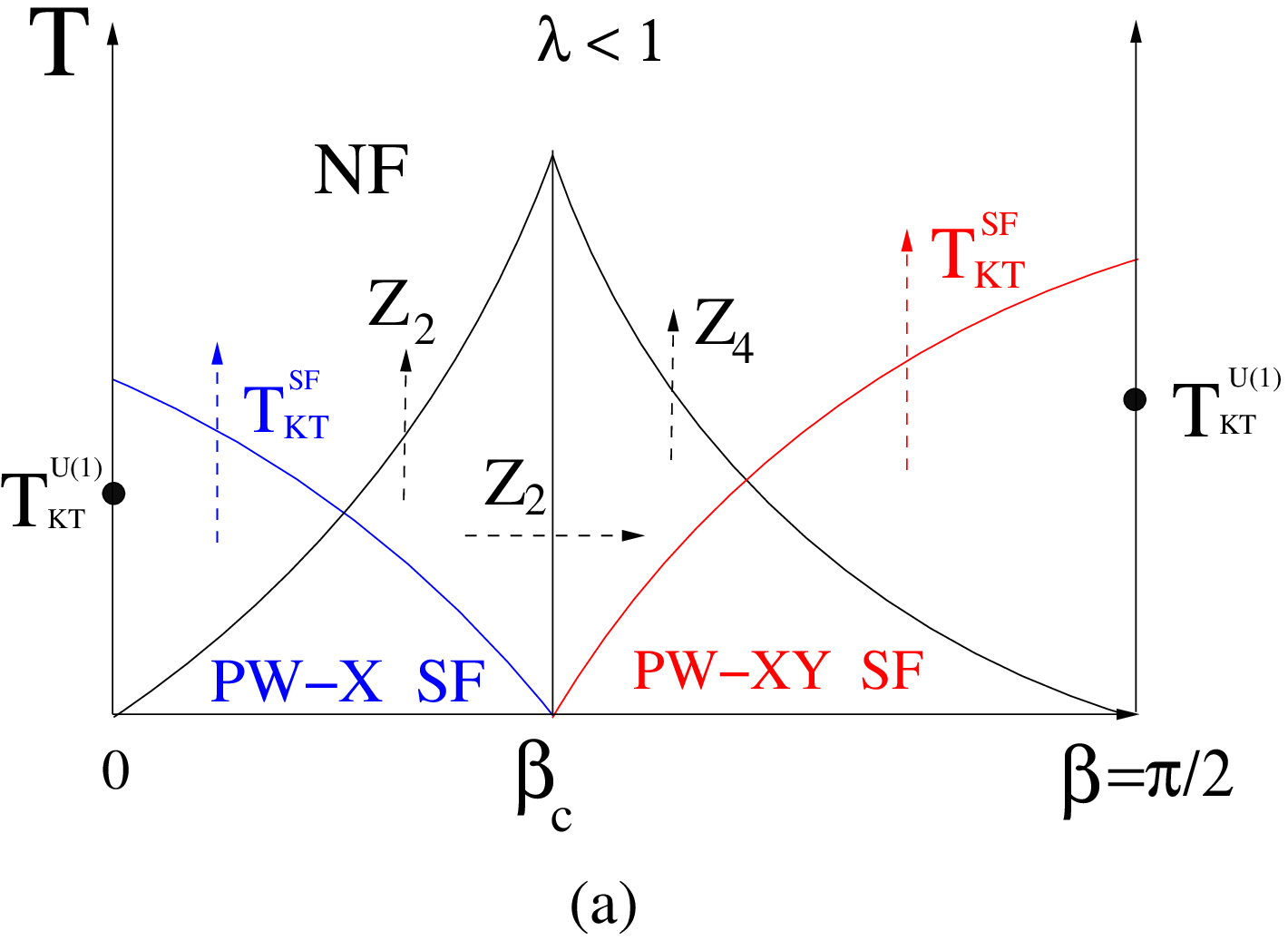}
    \includegraphics[width=0.225\textwidth]{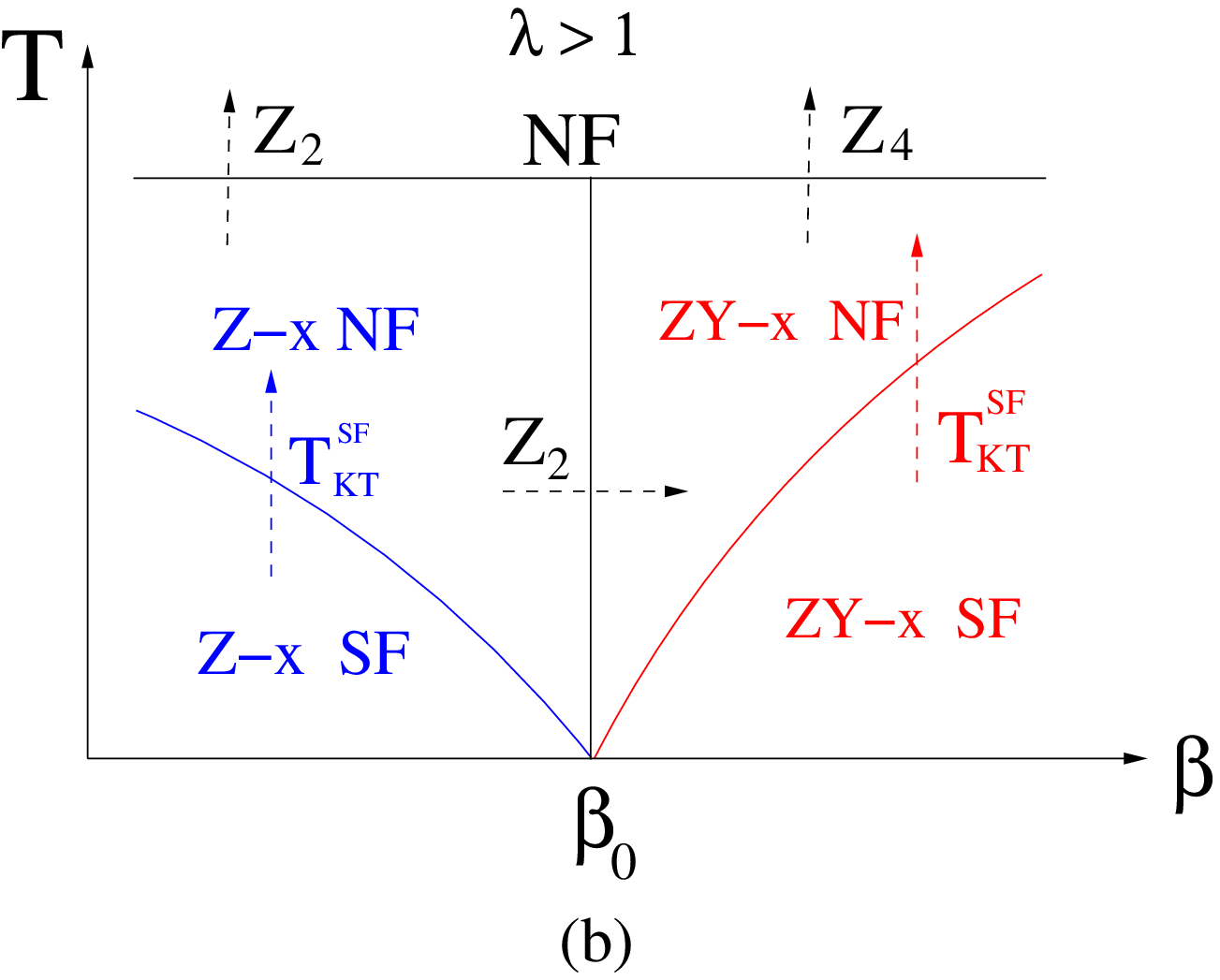}
    \caption{ The quantum phase transitions at $ T=0 $ are driven by the softening of SF along the $ y $ direction
    ( namely the bosonic Lifshitz transition with $ (z_x=1,z_y=2 ) $ ) in Eq.\ref{Gbetac}.
    The finite temperature phase transitions in (a) $ \lambda < 1  $.
      There are KT transitions at $ T^{SF}_{KT}  $ from the SF to a normal fluid (NF).
      There is also a Ising and $ Z_4 $ clock melting transition with
      $ T_2 \sim \Delta(\pi,0) $ and  $ T_4 \sim \Delta(\pi,2 k_0 ) $ above the PW-X SF and PW-XY SF respectively.
      Note that as $ \beta \rightarrow \beta^{+}_c, T_4 \sim \Delta(\pi,0) $ and $ \beta \rightarrow \pi/2^{-}, T_4 \sim \Delta(\pi,\pi) $.
      On the left and right axis, there is also a KT transition due to the $ U(1) $ symmetry breaking in the
      $\tilde{SU}(2) $ basis and $\tilde{\tilde{SU}}(2) $ basis respectively.
    (b) $ \lambda > 1 $. Note that as shown in Fig.1b, $ \beta_0 > \beta_c $.}
\label{sfdrive}
\end{figure}

  Now we briefly discuss the finite temperature phases and phase transitions above all the SFs in Fig.1b.
  Of course, due to the $ U(1)_c $ symmetry broken in the SF phase, there is always a KT transition $ T^{SF}_{KT} $ above all the SFs.
   Due to the correlated spin-bond orders of the SFs, there are also other interesting phase transitions associated  with the restorations of these spin-bond correlated orders at a finite temperature shown in Fig.\ref{rotondrive} and Fig.\ref{sfdrive}. The nature of these finite temperature transitions can be qualitatively determined by the degeneracy of the ground states:
  PW-X ( $ d=2 $ ), Z-x SF ( $ d=2 $ ), PW-XY ( $ d=4 $ ), ZY-x SF ( $ d=4 $ ).
  At $ \lambda=1 $, due to the breaking of $ U(1)_{soc} $, there is also a  KT transition
  $ T^{soc}_{KT}  $ in Fig.7.

\begin{figure}[!htb]
\centering
\includegraphics[width=3.75cm]{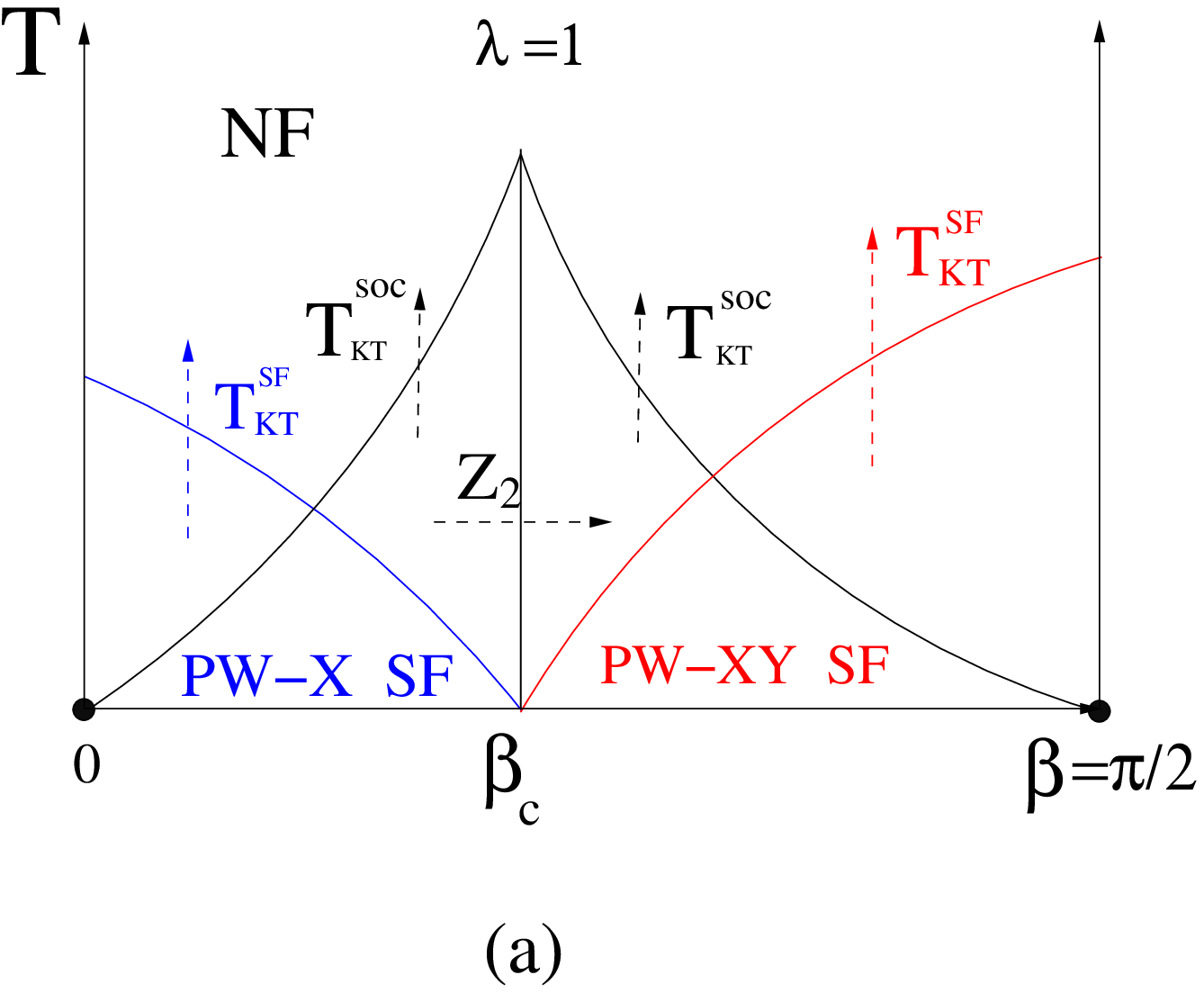}
\hspace{0.25cm}
\includegraphics[width=3.75cm]{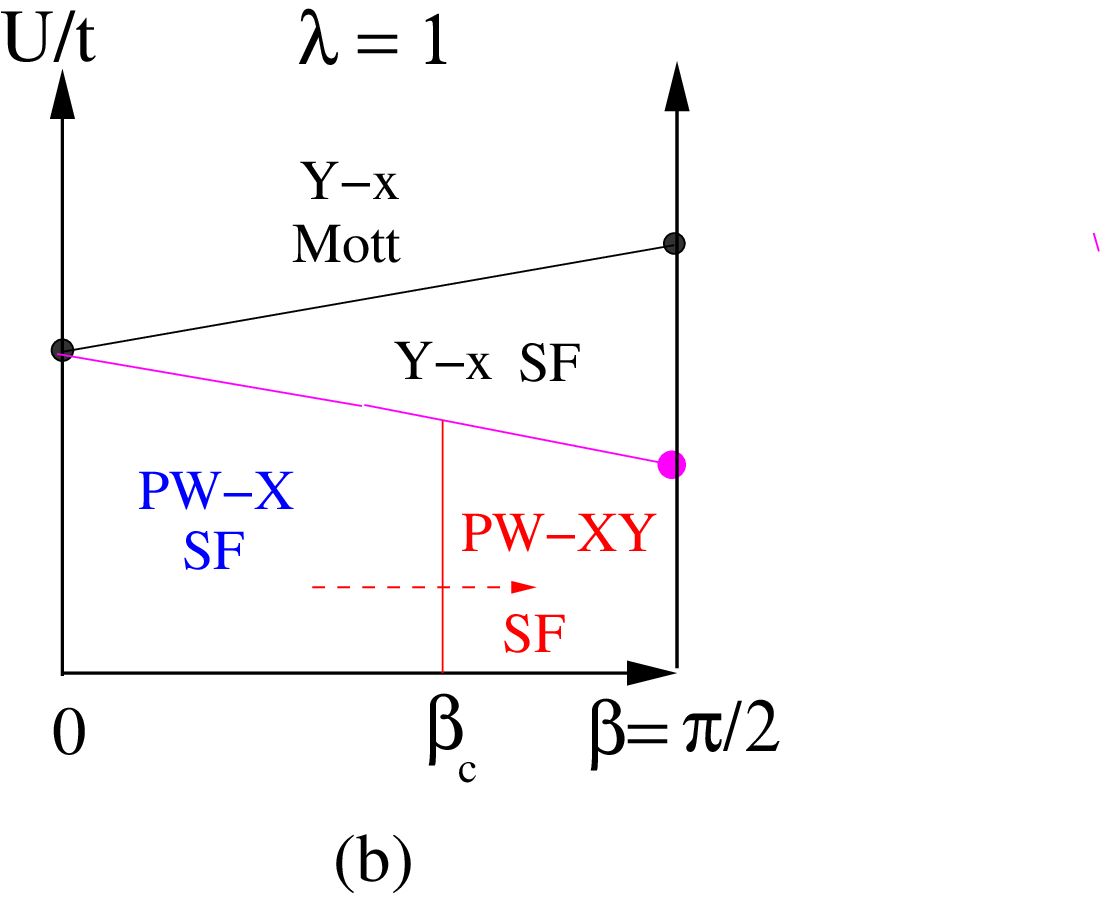}
    \caption{
   (a)   The finite temperature phase transitions at $ \lambda = 1  $.
    The finite temperature $ Z_2 $ and $ Z_4 $ transitions in the $ \lambda < 1 $ case in
    Fig.\ref{sfdrive}(a) are replaced by the KT transition in the spin sector.
    The left and right dot stands for the $\tilde{SU}(2) $ and  $\tilde{\tilde{SU}}(2) $  symmetry respectively.
    (b) The weak to strong coupling phase transitions at $ \lambda = 1 $.
    The two broken symmetries $ U(1)_c $ and $ U_s(1) $ may not be restored at the same time as the interaction increases.
    The Y-x SF phase breaks only the $ U(1)_{c} $,
    but not the $ U(1)_{soc} $  ( it could also be an insulating phase which breaks only the $ U(1)_{soc} $,  but not the $ U(1)_c $ ).
    If it is a  Y-x SF phase, then the transition across the black line is just the conventional SF-Mott transition.
    But the nature of the transition across the purple line remains to be explored.}
\label{sfdrive10}
\end{figure}

Now we briefly mention the transitions from the various SFs in Fig.1b at weak coupling to the
magnetic phases at the strong coupling  \cite{rh}.
At $ \lambda=1 $, at the weak coupling, the two SF states break both the $ U(1)_c $ order and also
the $ U(1)_{soc} $ symmetry, therefore supports two kinds of linear Goldstone modes shown in Fig.3b.
However, at the strong coupling, to the leading order of $ t^2/U $, it was shown \cite{rh} the
the Y-x state is the exact ground state not only free of frustrations, but also quantum fluctuations.
It is a gapped Mott state which keep both $ U(1) $ symmetry.
So the two SFs to the Y-x  state transition may go through the Y-x SF phase which still breaks the $ U(1)_{c} $ symmetry,
but restores the $ U(1)_{soc} $ symmetry shown in Fig.\ref{sfdrive10}b. It may also go through
an intermediate insulating phase which breaks the $ U(1)_{soc} $ symmetry, but not the $ U(1)_c $ symmetry.

{\bf 8. Implications and detections on cold atom experiments with Rashba SOC in optical lattices }

Due to the heating issues associated with the SOC generated by Raman laser scheme on alkali fermions, it would be difficult
to observe many body phenomena on alkali fermions with the Raman scheme.
However, optical lattice clock schemes \cite{clock} have been successfully implemented \cite{clock1,clock2}
to generate SOC for $^{87} Sr $ and $ ^{137} Yb $. this newly developed scheme has the advantage to suppress
the heating issue suffered in the Raman scheme.
It can also probe the interplay between the interactions and the SOC easily.


   The TOF image after a time $ t $ is given by \cite{blochrmp}:
 \begin{equation}
     n( \vec{x} )= ( M/\hbar t )^3 f(\vec{k}) G(\vec{k})
 \end{equation}
   where $ \vec{k}= M \vec{x}/\hbar t $,  $ f(\vec{k})= | w(\vec{k}) |^2 $ is the form factor due to the
   Wannier state of the lowest Bloch band of the optical lattice and
   $ G(\vec{k}) = \frac{1}{N_s} \sum_{i,j} e^{- \vec{k} \cdot ( \vec{r}_i- \vec{r}_j ) }
    \langle \Psi^{\dagger}_i \Psi_j \rangle $ is the equal time boson structure factor.
   For small condensate depletion in the weak coupling limit $ U/t \ll 1 $,
   $ \langle \Psi^{\dagger}_i \Psi_j \rangle \sim \langle \Psi^{\dagger}_{0i} \Psi_{0j}  \rangle$
   where $ \Psi^{\dagger}_{0i} $ is the condensate wavefunction Eq.\ref{twonodes} at the mean field level.
   So the TOF can detect the quantum ground state wavefunction directly.

 All the SF phases and their excitations across the whole BZ in Fig.1b,
 the transitions driven by the softening of the SF Goldstone mode or the roton touch down in
 Fig.\ref{twoexp} can be precisely determined by various experimental techniques
 such as dynamic or elastic, energy or momentum resolved, longitudinal or
 transverse Bragg spectroscopies \cite{braggbog,braggangle,braggeng,braggsingle,becbragg,bragg1,bragg2},
 specific heat measurements \cite{heat1,heat2}  and {\sl In-Situ} measurements \cite{dosexp}.

{\bf  DISCUSSIONS }

 In fact, all the quantum phases and phase transitions in Fig.1 can be
 summarized as in the following global picture:
 The SOC parameter $ \beta $ splits the phase diagram into a
 1D SOC regime where there are only two minima at $ (0, \pm \pi/2 ) $ ㄗ which are independent of the SOC parameter $ \beta $ )
 and a 2D SOC regime where there are 4 minima  at $ ( \pm \pi/2, \pm k_0) $ ㄗ which depend on the SOC parameter $ \beta $ ).
 The two regimes are connected by the bosonic Lifshitz transition driven by the softening of SF Goldstone mode
 with the dynamic exponent $ (z_x=1, z_y=2) $ and the accompanying $ Z_2 $ transition breaking the $ {\cal P}_x $ symmetry in the spin sector.
 The spin anisotropy parameter $ \lambda $ splits each regime into two: below $ \lambda=1 $ the states are plane wave (PW) states,
 above $ \lambda=1 $, the states are stripe ones breaking translational symmetries of the lattice.
 They are connected by second order transitions driven by the roton touchdowns.
 The Tetra-critical point in the  $ \lambda=1 $ line can be considered as the bosonic version of
 the fermionic topological transitions also connecting the two
 Abelian points with flux 0 and $ \pi $ driven by the collisions of 2d Dirac fermions \cite{tqpt} or 3d Weyl fermions \cite{weylscaling}.
 The extension to generic Rashba SOC parameter $ ( \alpha, \beta ) $ which tunes the strengths of the frustrations will be presented
 in a separate publication \cite{junsen}. Of course, any $ \alpha \neq \pi/2 $ breaks  the $ U(1)_{soc} $ symmetry explicitly.

 As elucidated in \cite{qahsf}, there is just one sign difference on the down spin hopping between the the Rashba SOC model Eq.\ref{intlambda}
and the quantum anomalous Hall (QAH) model \cite{2dsocbec}. But this sign difference makes the two models show dramatically different phenomena.
In the QAH model, the system has the $ [C_4 \times C_4]_D $ symmetry, at $ h=0 $, it has an extra anti-unitary Reflection $ Z_2 $ symmetry.
The transition driven by roton dropping is a first order one, the quantum Lifshitz transition driven by the softening of the SF Goldstone mode
is a second order one. The two transitions meet at a topological Tri-critical (TT) point which separates
the $ N=2 $ XY-AFM from the $ N=4 $ XY-AFM at $ h=0 $ line. Both phases are gapped in the spin sector and break the $ [C_4 \times C_4]_D $ symmetry, but keeps the Reflection symmetry, so can only be distinguished by the topology of the BEC condensation momenta
on the two sides of the transition instead of any symmetry principles. At the TT, the condensation momenta are along the two crossing lines $ k_x= \pm k_y $ and the  dispersion also becomes flat along the two lines.
In the Rashba SOC model, the system has the three unitary $ {\cal P}_x, {\cal P}_y, {\cal P}_z $ symmetries at any $ \lambda $.  It has an extra $ U(1)_{soc} $ symmetry at $ \lambda=1 $. The transition driven by roton touchdown is a second order one, the quantum Lifshitz transition driven
by the softening of the SF Goldstone mode is also a second order one. The two transitions meet at a Tetra-critical point at $ \beta_c $
which separates the PW-X phase from the PW-XY phase at the $ \lambda=1 $ line.
Both phases support the gapless linear spin Goldstone mode due to the $ U(1)_{soc} $ breaking,
the PW-X phase keeps the $ {\cal P}_x $ symmetry, but
the PW-XY phase breaks the $ {\cal P}_x $ symmetry. Both break the  $ {\cal P}_y, {\cal P}_z $ symmetry,
so there is an accompanying $ {\cal P}_x $  symmetry breaking across the T critical point.
Across the T point at $ \beta_c $  ( Fig.1a), there is just $ N=2 $ BEC condensation momenta split into $ N=4 $ BEC condensation momenta
instead having any continuous distributions of BEC condensation momenta or flat directions in the dispersion relation, so the transition is just a conventional symmetry breaking transition instead of a topological one.      The Rashba model in a Zeeman field at $ \lambda=1 $ will be presented in a separate publication \cite{junsen}.
Of course, any Zeeman field breaks explicitly the $ U(1)_{soc} $ symmetry.
The Rashba SOC (QAH) model is topological trivial ( non-trivial ).
So this work is complementary to that  in the QAH model studied in \cite{qahsf}.
        The results to be achieved may shed lights on each other.
        It remains interesting to investigate possible connections between
the many body phenomena of weakly interacting spinor bosons and the topology of the underlying non-interacting energy bands.


 The PW-X to Z-x ( or PW-XY to ZY-x )  phase transition as shown in Fig.\ref{twoexp} can also be contrasted with a superfluid (SF)
 to a supersolid (SS) transition  of a single component hard core boson
 in a triangular lattice slightly away from $1/3$ ( or $ 2/3 $ ) filling, also driven by a roton touchdown
 \cite{gan,ss1,ss2,ss3,tri,trinnn,five,yan}. The main difference here is:
 In the former, the roton touchdown is tuned by the spin anisotropic parameter $ \lambda $ inside a superfluid phase, at any fillings within a weak coupling limit. The translational symmetry breaking in Z-x ( or ZY-x ) comes from the spin-orbit coupling.
 While in the latter,  the roton touchdown is tuned by $ t/V_1 $ where $ V_1 $ is the nearest neighbor (NN) interaction,
 near the SF to Mott transition at $1/3$ ( or $ 2/3 $ ) filling.
 The solid component comes from the underlying Mott phase at $1/3$ ( or $ 2/3 $ ) filling. The SF component
 comes from the interstitials ( or vacancies ) relative to the Mott phase.

{\bf  Methods }

{\bf 1. The spin-orbital structure and the degeneracy of the Z-x and ZY-x state. }

When $\beta>\beta_c$, as shown in Fig.M1a, the spectrum has 4 minima at
$\mathbf{K}_1=(\frac{\pi}{2},k_0)$,
$\mathbf{K}_2=(-\frac{\pi}{2},k_0)$,
$\mathbf{K}_3=(-\frac{\pi}{2},-k_0)$,
$\mathbf{K}_4=(\frac{\pi}{2},-k_0)$,
where $k_0=\arcsin\sqrt{\sin^2\beta-\cot^2\beta}$.
The corresponding spinor is
\begin{align}
    \chi_i
    =\frac{1}{\sqrt{2}}
    \begin{pmatrix}
    1\\
    -e^{i\phi_i}\\
    \end{pmatrix},\quad i=1,2,3,4
\label{spinor0}
\end{align}
 where $\phi_1=\text{arcsec}(\sin\beta\sqrt{1+\sin^2\beta})=\phi_0$,
$\phi_2=\pi-\phi_0$,
$\phi_3=\pi+\phi_0$,
$\phi_4=-\phi_0$.

 The 4-fold degenerate ZY-x state can be written as
\begin{align}
    \Psi_{1\pm2}^\text{ZY-x}
	=\frac{1}{\sqrt{2}}e^{-i\mathbf{K}_1\cdot\mathbf{r}_i}\chi_1
	\pm
	\frac{1}{\sqrt{2}}e^{i\phi_0}e^{-i\mathbf{K}_2\cdot\mathbf{r}_i}\chi_2, \nonumber   \\
    \Psi_{3\pm4}^\text{ZY-x}
	=\frac{1}{\sqrt{2}}e^{-i\mathbf{K}_3\cdot\mathbf{r}_i}\chi_3
	\pm
	\frac{1}{\sqrt{2}}e^{i\phi_0}e^{-i\mathbf{K}_4\cdot\mathbf{r}_i}\chi_4
\end{align}
 whose spin-orbital structure can be easily evaluated as
\begin{align}
    \langle\Psi_{1\pm2}^\text{ZY-x}|\vec{\sigma}|\Psi_{1\pm2}^\text{ZY-x}\rangle
	=(0,-\sin\phi_0,\pm(-1)^{i_x}\cos\phi_0)       \nonumber   \\
    \langle\Psi_{3\pm4}^\text{ZY-x}|\vec{\sigma}|\Psi_{3\pm4}^\text{ZY-x}\rangle
	=(0,+\sin\phi_0,\pm(-1)^{i_x}\cos\phi_0)
\label{ZYx}
\end{align}
which are shown in the following figure.
\begin{figure}[!htb]
    \includegraphics[width=0.4\linewidth]{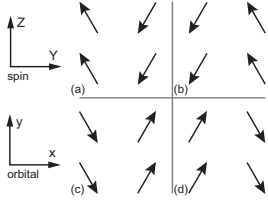}
	\caption{ The spin-orbital structure of the 4-fold degenerate ZY-x state at $\lambda>1$ and $\beta>\beta_c$.}
\end{figure}

    It is easy to see that the state differs by translation by one lattice site along the row,
    but reverses the Y component of the spin along the column.
    If setting $ k_0=0, \phi_0=0 $, the Y component vanishing,
    the 4-fold degenerate ZY-x reduces to the  2-fold degenerate Z-x state in
    the Fig.M1b.

{\bf 2. The kinetic energy and currents in all the 4 phases in Fig.1. }

   Using the method in \cite{yan}, we will evaluate the conserved density current in all the 4 phases in Fig.1.
   In the context of 2d charge-vortex duality \cite{yan},
   the vortex currents in the dual lattice gives the boson densities in the direct lattice.

  We can write the kinetic energy in Eq.2 as:
\begin{align}
    \mathcal{H}_\text{hop}
	=\sum_{i,\mu}(\psi_i^\dagger H_\mu \psi_{i+\mu}+h.c.),
\end{align}
  where $ H_{x}=-t e^{i\frac{\pi}{2}\sigma_x}
	     =-it\sigma_x,
	H_{y}=-t e^{i\beta\sigma_y}
	     =-t(\cos\beta~\sigma_0+i\sin\beta~\sigma_y) $.

    Along a given bond $ (i, i+ \mu) $:
\begin{equation}
    \psi_i^\dagger H_\mu \psi_{i+\mu}  =K-i I
\end{equation}
  where $ K $ is the kinetic energy and $ I $ is the current flowing along the bond \cite{yan}.

  As shown in Fig.1, for $\lambda\leq1$, the ground-state is a PW state
\begin{align}
    \psi_i=e^{i\mathbf{K}_1 \cdot\mathbf{r}_i}\chi_1,
\end{align}
where, as listed in Eq.\ref{spinor0}, the $k_0$ and $\phi_0$ take the piecewise form
\begin{align}
    k_0=\begin{cases}
	0,  &  \beta<\beta_c   \\
	\arcsin\sqrt{\sin^2\beta-\cot^2\beta}, &  \beta>\beta_c   \\
	\end{cases}
\end{align}
and
\begin{align}
    \phi_0	=\begin{cases}
	0,  & \beta<\beta_c   \\
	\text{arcsec}(\sin\beta\sqrt{1+\sin^2\beta}), & \beta>\beta_c   \\
	\end{cases}
\end{align}

  By using the relations between $k_0$ and $\phi_0$
\begin{align}
    \tan k_0=\tan\beta\sin\phi_0,
    \cos\phi_0=\frac{1}{\sqrt{1+\sin^2\beta\sin^2 k_0}}
\end{align}

  We find
\begin{align}
	\psi_i^\dagger H_x \psi_{i+x}
	=\begin{cases}
	    -t, &\beta<\beta_c\\
	    \frac{-t}{\sin\beta\sqrt{1+\sin^2\beta}} \leq -t, &\beta>\beta_c
	\end{cases}
\end{align}
and
\begin{align}
    \psi_i^\dagger H_y \psi_{i+y}
	=\begin{cases}
	    -t\cos\beta, &\beta<\beta_c\\
	    \frac{-t\sin\beta}{\sqrt{1+\sin^2\beta}} \leq -t\cos\beta, &\beta>\beta_c
	\end{cases}
\end{align}
   So there are no currents in both the PW-X and PW-XY states.
   It is easy to show that the two equalities hold at the quantum Lifshitz transition point $ \beta_c $.

   For $\lambda>1$, the 4 fold degenerate ground-state were written in Eq.\ref{ZYx}.
   We find the results are the same as  the $\lambda\leq 1$ case.
   So the kinetic energies stay the same across the second order transition at $ \lambda=1 $.

{\bf 3. The gap and Goldstone mode generated by the order from disorder mechanism }

 In this method section, we provide a unified scheme to compute the mass generation of the roton mode at $\lambda<1$
 and the computation of spectrum of the Goldstone mode at $\lambda=1$ by the order-from-disorder mechanism.
 The relations between the parametrization in the PW-X at $ \lambda < 1 $ and $ \lambda=1 $ has been established in
 Eq.\ref{exch}. Here, we may just use the order from disorder variables in the $ \lambda < 1 $ case.
 Since $B(\lambda)$ in Eq.\ref{A2B2} and Eq.\ref{B2} is continuous at $\lambda=1$, so we only need focus on $\lambda<1$ case
 and then take $ \lambda \rightarrow 1^{-} $ limit.
 We need to establish relation between the $\phi$ variable in Eq.\ref{A2B2} in the order from disorder analysis
 and the original Bose fields in Eq.\ref{intlambda}.

 The Bose condensation Eq.\ref{twonodes} inspires us to parameterize the most general Bose field in the polar-like coordinate system as
\begin{align}
    \Psi_i
	=e^{i\chi}\sqrt{\rho}
	\left[
	c_1\eta_1
	e^{i\mathbf{K}\cdot\mathbf{r}_i}
	+
	c_2\eta_2
	e^{-i\mathbf{K}\cdot\mathbf{r}_i}
	\right]
\end{align}
where one can parameterize the two coefficients $c_{1,2}$  in the most general forms as in Eq.\ref{c1c2}
\begin{align}
    c_1=[e^{i\phi/2}\cos(\theta/2)+e^{-i\phi/2}\sin(\theta/2)]/\sqrt{2},  \nonumber  \\
    c_2=[e^{i\phi/2}\cos(\theta/2)-e^{-i\phi/2}\sin(\theta/2)]/\sqrt{2}
\end{align}
and the two components spinors $\eta_{1,2}$ as:
\begin{align}
	\eta_1=
	\begin{pmatrix}
	    e^{+i\phi_1/2}\cos(\theta_1/2)\\
	    e^{-i\phi_1/2}\sin(\theta_1/2)\\
	\end{pmatrix},
	\quad
	\eta_2=
	\begin{pmatrix}
	    e^{+i\phi_2/2}\cos(\theta_2/2)\\
	    e^{-i\phi_2/2}\sin(\theta_2/2)\\
	\end{pmatrix}
\end{align}
Thus we can write $\Psi_i=\Psi_i(\rho,\chi,\theta,\phi,\theta_1,\phi_1,\theta_2,\phi_2)$.

Since we already found the quantum ground state is the PW-X state
which corresponds to the saddle point values:
\begin{align}
   \Psi_{i,0} =(\rho_0,0,\pi/2,0,-\pi/2,0,\pi/2,0)\>.
\end{align}
   One can write down the quantum fluctuations around the saddle point as:
\begin{align}
    \Psi_i=\Psi_{i,0} + ( \delta\rho,\chi,\delta\theta,\phi,
\delta\theta_1,\phi_1,\delta\theta_1,\phi_2 )
\end{align}

Now, we can separate the Bose field into the condensation part plus the quantum fluctuation part
in the polar coordinate system.
\begin{align}
    \Psi_i
	& =\sqrt{\frac{\rho_0}{2}}
	\begin{pmatrix}
	    1\\
	   -1\\
	\end{pmatrix}
	e^{i\mathbf{K}\cdot\mathbf{r}_i}    \nonumber  \\
	& +
	\sqrt{\frac{\rho_0}{8}}
	\left[
	(\frac{\delta\rho}{\rho_0}+i\chi)
	\begin{pmatrix}
	    1\\
	    -1\\
	\end{pmatrix}
	+
	(\delta\theta_1+i\phi_1)
	\begin{pmatrix}
	    1\\
	    1\\
	\end{pmatrix}
	\right]
	e^{i\mathbf{K}\cdot\mathbf{r}_i}     \nonumber    \\
   & + \sqrt{\frac{\rho_0}{8}}
	(-\delta\theta+i\phi)
	\begin{pmatrix}
	    1\\
	    1\\
	\end{pmatrix}
	e^{-i\mathbf{K}\cdot\mathbf{r}_i}
\label{poldecomp}
\end{align}
  where the third line explicitly contains the order from disorder variable $ \phi $.

  As written down in the previous sections, in the original Cartesian coordinate, we have used:
\begin{align}
    \Psi_i=\sqrt{\frac{N_0}{2}}
	\begin{pmatrix}
	    1\\
	   -1\\
	\end{pmatrix}
	e^{i\mathbf{K}\cdot\mathbf{r}_i}
	+
	\begin{pmatrix}
	    \delta\psi_{i\uparrow}\\
	    \delta\psi_{i\downarrow}\\
	\end{pmatrix}
\label{cardecomp}
\end{align}

   After comparing both the $ e^{i\mathbf{K}\cdot\mathbf{r}_i} $ and the  $ e^{-i\mathbf{K}\cdot\mathbf{r}_i} $
   components in the two equations  Eq.\ref{poldecomp} and Eq.\ref{cardecomp}, one can express the order from disorder variable $ \phi $
   in terms of the original Bose field:
\begin{align}
    \phi(\mathbf{q})
    & =-\frac{i}{\sqrt{2\rho_0}}
    [\delta\psi_{-\mathbf{K}+\mathbf{q},\uparrow}
    +\delta\psi_{-\mathbf{K}+\mathbf{q},\downarrow}   \nonumber  \\
    & -\delta\bar{\psi}_{\mathbf{K}-\mathbf{q},\uparrow}
      -\delta\bar{\psi}_{\mathbf{K}-\mathbf{q},\downarrow}]
\end{align}
   where as expected, only the quantum fluctuations near $ -\mathbf{K} $ appear in the relation.

  Thus we can express the quantum correction coming from order-from-disorder mechanism in terms of
  the original Bose fields as
\begin{align}
	\delta H
	=\sum_i\frac{B}{2}\phi_i^2
	=\sum_q\frac{B}{2}\phi_q\phi_{-q}
\label{corr}
\end{align}
  which, we must stress, only holds near $ -\mathbf{K} $, so it will not affect the Goldstone mode near $ \mathbf{K} $.

Finally, after combining with  $ H_2 $ in Eq.\ref{h2},
we arrive at the Order from disorder (OFD) corrected Hamiltonian:
\begin{align}
    H_\text{OFD}=H^{(2)}+\delta H
	=\frac{1}{2}\sum_{q}\Psi_q^\dagger (M+\delta M) \Psi_q
\label{total}
\end{align}
   where the $ 4 \times 4 $ matrix $M$ was already obtained from the PW-X calculation leading to Eq.\ref{h2}
   and the $\delta M$ in Eq.\ref{corr} can be written as in a $ 4 \times 4 $ matrix form:
\begin{align}
    \delta M=
	\frac{B}{2n_0}
	\begin{pmatrix}
		1 &1 &-1 &-1\\
		1 &1 &-1 &-1\\
		-1&-1&1  &1\\
		-1&-1&1  &1\\
	\end{pmatrix}
\end{align}

 We diagonalize the order from disorder corrected Hamiltonian Eq.\ref{total}
 by a $ 4 \times 4 $ Bogoliubov transformation:
\begin{align}
    H_\text{OFD}=\sum^{2}_{l=1} \sum_{q \in BZ }\omega_{l,q}
	\left(\beta_{l,q}^\dagger\beta_{l,q}+\frac{1}{2}\right)
\end{align}
where $\omega_{1,q}<\omega_{2,q}$. So one only need to focus
on $ \omega_1 $.

From the complete form of the dispersion $ \omega_1 ( -\vec{K} + \vec{q} ) $,
we can extract the long wavelength limit of the roton mode as
\begin{align}
    \omega_R(\mathbf{q})
	= \sqrt{\Delta_R + B_R
	 [ q_x^2	+(\cos\beta-C\sin\beta^2 ) q_y^2]}
\label{gapdisp}
\end{align}
   where
   $\Delta_R=\omega_R(\mathbf{q}=0)=\sqrt{2BU(1-\lambda)}$ is the roton gap in Eq.\ref{gap}
   generated by the order-from-disorder mechanism. The two coefficients
   $ B_R=2Bt/n_0+n_0Ut(1-\lambda)$ and
   $ C=1+\frac{BU[4t(1-3\lambda)+2B(1-\lambda)/n_0+n_0U(1-\lambda)^2]}
   {[2B/n_0+n_0U(1-\lambda)][8t^2+2n_0Ut(1+\lambda)-BU(1-\lambda)]}<1 $
   represent the corrections to the Roton dispersion.
   Obviously, setting $ B=0 $, Eq.\ref{gapdisp} recovers the second equation in Eq.\ref{GR}.


   Of course, due to the momentum separation as stressed below Eq.\ref{corr},
   the superfluid Goldstone mode $ \omega_1 ( \vec{q} ) $ dictated by the $ U_c(1) $ symmetry breaking
   remains the same as listed in Eq.\ref{GR} as

   Going beyond Eq.\ref{gap}, Eq.\ref{gapdisp} not only contains the roton gap $\Delta_R $ in Eq.\ref{gap},
   but also the correction to the  dispersion relation due to the order from disorder formalism.
   To the best of our knowledge, this is the first " order from disorder " calculation of the correction to the
   dispersion relation beyond a gap calculation.

   By taking $\lambda \to1 ^-$ limit in Eq.\ref{gapdisp},
   the roton gap  vanishes and recovers the linear dispersion in Eq.\ref{rotonl}.

{\bf Acknowledgements }

  F. Sun and J. Y acknowledge AFOSR FA9550-16-1-0412 for supports.
  The work at KITP was supported by NSF PHY11-25915.
  J.W and Y.D thank the support by NSFC under Grant No. 11625522 and the Ministry of Science and
  Technology of China (under grants 2016YFA0301604).


\begin{thebibliography}{99}
\bibitem{notation} Here we still use the same notation used in \cite{rh}.
The PW-X means the plane wave state with the spin polarization along the X direction.
The PW-XY means the plane wave state with the spin polarization along the XY direction.
In the Z-x ( or ZY-x ) phase, the first Capital letter indicates
the spin polarization, the second small letter indicates the orbital order.
\bibitem{analogy} It maybe common that the exponents are different on two sides of topological phase transitions \cite{tqpt,weylscaling},
but it seems this is a first example which happens across a QPT. It is a new phenomenum due to the frustrations from the SOC.
\bibitem{subtract}  Because the bosons can condense at one of the two minima in Fig.1a1. In the following, the momentum
is measured relative to this condensation point.
\bibitem{notesy}  Both Abelian points contain $ \sum_i  (-1)^{i_x}S_i^y $ as the conserved quantity.
 In fact, it remains the conserved quantity generating the $ U(1)_{soc} $ symmetry connecting the two Abelian points
 along $ \lambda=1 $ line in Fig.1
\end{thebibliography}
\end{document}